\documentclass[final,12pt]{elsarticle}

\usepackage[export]{adjustbox}
\usepackage{amssymb}
\usepackage{amsthm}
\usepackage{upgreek}
\usepackage{hyperref}
\hypersetup{
	colorlinks = true,
	urlcolor   = blue,
	citecolor  = black,
}
\usepackage{float}
\usepackage{multirow}

\newcommand{\RvMJ}[1]{\textcolor{black}{#1}}
\begin{document}

\begin{frontmatter}



\title{Stability and accuracy of the weakly compressible SPH with particle regularization techniques}


\author[inst1]{Mojtaba Jandaghian}
\ead{mojtaba.jandaghian@polymtl.ca}
\author[inst1]{Herman Musumari Siaben}
\ead{herman.musumari-siaben@polymtl.ca}
\author[inst1,inst2]{Ahmad Shakibaeinia\corref{cor1}}
\ead{ahmad.shakibaeinia@polymtl.ca}

\affiliation[inst1]{organization={Department of civil, geological, and mining engineering},
            addressline={Polytechnique Montreal}, 
            city={Montreal},
            country={Canada}}

	
\affiliation[inst2]{organization={Canada research chair in Computational Hydrosystems}}  
\cortext[cor1]{Corresponding author}
\journal{European Journal of Mechanics - B/Fluids}
\begin{abstract}
This paper proposes and validates two new particle regularization techniques for the Smoothed Particle Hydrodynamics (SPH) numerical method to improve its stability and accuracy for free surface flow simulations. We introduce a general form of the Dynamic pair-wise Particle Collision (DPC) regularization technique that we recently proposed in the context of the Moving Particle Semi-implicit (MPS) method in \cite{Jandaghian2021_JCP}. The DPC coupled with the standard Particle Shifting (PS) technique has given rise to a hybrid approach that we propose to alleviate particle clustering issues in the free-surface and splashed regions. We validate the proposed techniques to four benchmark cases: (i) the oscillating droplet, (ii) the two-dimensional water dam-break, (iii) the two-dimensional water sloshing, and (iv) the three-dimensional water dam break against a rigid obstacle. We evaluate their impacts on the stability,  accuracy and the conservation properties of the test cases. The qualitative and quantitative analysis of the results shows that despite its simplicity, the DPC technique is more effective in reducing the spatial disorder and capturing the impact events compared with the standard and the newly improved hybrid PS methods. Although the hybrid PS technique improves particle distribution at the free surface, it still suffers from the inconsistent implementation of the PS equation which unphysically increases the fluid volume and violates the conservation of potential energy in the long-term simulations. Overall, the conservative DPC algorithm proves to be a simple and efficient alternative regularization technique for simulating such highly dynamic free-surface flows.
\end{abstract}

\begin{graphicalabstract}
\includegraphics[width=\textwidth]{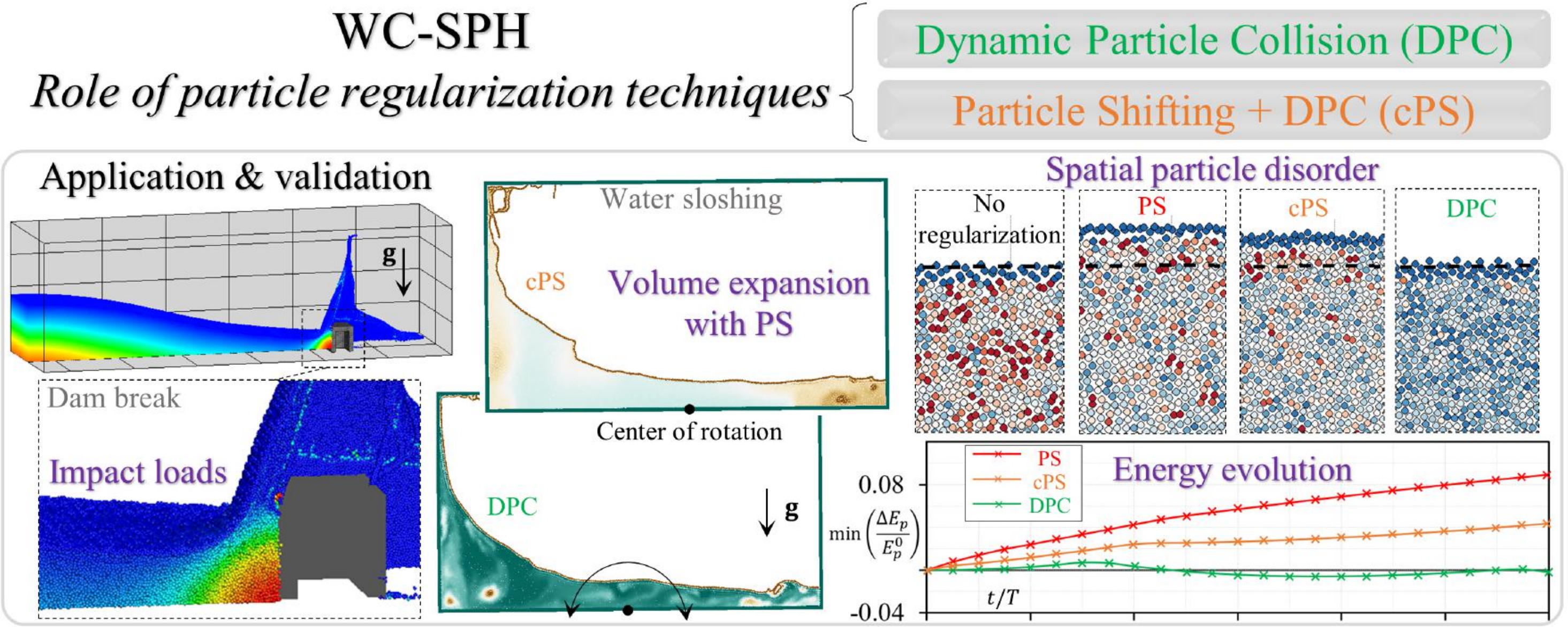}
\end{graphicalabstract}

\begin{highlights}
\item Role of particle regularization techniques in the stability and accuracy of SPH
\item A conservative and effective particle collision technique for SPH
\item Dynamic particle collision technique improves the spatial particle disorder
\item A hybrid particle shifting approach reduces the particle clustering issues in the free-surface regions
\end{highlights}


\end{frontmatter}


\section{Introduction}
\label{sec:Intro}
Continuum-based mesh-free particle methods, such as the Smoothed Particle Hydrodynamics (SPH) \cite{Gingold&Monaghan1977} and Moving Particle Semi-implicit (MPS) \cite{Koshizuka1996} methods, have remarkable capabilities for simulating highly dynamic free-surface and interfacial flows (e.g., for solving multiphysics problems in industry \cite{Shadloo2016, Li2020} and in hydro-environmental fields \cite{Jandaghian2021_AWR, Luo2021}). Nevertheless, due to their Lagrangian nature, they are prone to non-uniform particle distribution and unphysical spurious pressure noises that compromise their accuracy and stability. To deal with these issues some of the past studies have proposed higher-order approximation operators (e.g., \cite{Jandaghian2021_AWR,Khayyer2012}) or added the numerical diffusion terms to the continuity \cite{Molteni&Colagrossi2009, Jandaghian2020} and momentum \cite{Monaghan&Gingold1983} equations. But even with these improvements, the particle methods can still suffer from particle distribution anomalies. Therefore, another group of enhancement techniques works on the regularization of the particle distribution, among which, the particle shifting (PS) technique \cite{Xu2009, Lind2012}, the pair-wise particle collision (PC) method \cite{Lee2011,Shakibaeinia_cmame_2012}, and the transport-velocity formulation \cite{Adami2013,Zhang2017} are common approaches developed for SPH and MPS methods. 

In the SPH framework, Xu et al. \cite{Xu2009} proposed the PS method, which moves particles to the area less populated, avoiding therefore inter-particle penetrations. Lind et al. \cite{Lind2012} and Skillen et al. \cite{Skillen2013} adapted PS to free-surface flows involving body-water impacts. Further, Khayyer et al. \cite{Khayyer2017} proposed the optimized PS method which cancels the normal component of PS at the free-surface and its neighborhood. The pair-wise particle collision technique, which is based on the collision of physical solid or gas particles, was first time formulated within the context of the incompressible MPS by Lee et al \cite{Lee2011} (for free-surface flows) and for the weakly compressible MPS by Shakibaeinia and Jin \cite{Shakibaeinia_cmame_2012} (for multiphase flows).

Without special treatments of free-surface particles and considering the effects of the shifting transport-velocity, the PS algorithms (particularly implemented in the weakly compressible models) have shown to affect the mechanical behavior of the complex flows where the potential energy is dominant and several breaking events occur \cite{Sun2019,Krimi2020}. By applying PS, the potential energy of the system increases as observed by unphysical volume expansions discussed in \cite{Jandaghian2021_JCP} and \cite{Krimi2020}. In SPH, Sun et al. \cite{Sun2019} derived additional diffusive terms (added to the continuity and momentum equations) and implemented restricted boundary conditions to resolve the excessive potential energy due to continuous activation of PS. In MPS, Jandaghian et al. \cite{Jandaghian2021_JCP} proposed a consistent form of a corrected PS method by including the additional transport-velocity terms and implementing special boundary treatments (for large flow curvatures and by excluding the solid boundary particles). Sun et al. \cite{Sun2019} and Jandaghian et al. \cite{Jandaghian2021_JCP} showed that their developed PS algorithms, applied to weakly compressible particles methods, renders the latter free from unphysical volume expansions and numerical results divergence. 

\RvMJ{More recently, Lyu and Sun \cite{Hong-Guan2021} enhanced the SPH method by introducing a corrective cohesive force to tackle the volume-non-conservative issue of the original shifting equation. Antuono et al. \cite{Antuono2020} considered the consistent form of PS within an arbitrary Lagrangian-Eulerian SPH framework (presented by Oger et al. \cite{Oger2016}) improving the stability of the developed model by adopting artificial diffusion terms. Antuono et al. \cite{Antuono2021} coupled a Large-Eddy Simulation SPH (LES-SPH) method with the consistent PS and the Tensile Instability Control (TIC) scheme (proposed by Sun et al. \cite{Sun2018_MR-dplus}) for simulating high Reynolds number problems. Also, Lyu et al. \cite{Lyu2021} appended the PS formulation to the TIC method to surmount the tensile instability issue due to strong negative pressures in fluid-structure interactions with SPH. Marrone et al. \cite{Marrone2021} evaluated the dissipation mechanism in sloshing problems through the developed LES-SPH model with the enhanced particle stabilization techniques. Other research works, e.g., \cite{Zhang2021, Yang2022, Sato2021}, have successfully validated SPH models supplied with PS and artificial diffusion terms to study violent sloshing problems and dam-breaking flows with complex breaking waves. Further, Wen et al. \cite{Wen2021} adopted the multiphase particle collision model to ensure numerical stability of their developed incompressible MPS method for simulating violent multiphase flows.}

Jandaghian et al. \cite{Jandaghian2021_JCP} developed a new version of the particle collision technique within their enhanced weakly compressible MPS method for simulating violent free-surface flows. They validated that the proposed algorithm, denoted as the Dynamic pair-wise Particle Collision (DPC) technique, captures the impact events of such complex flows and eliminates particle pairing instability over the fluid domain. Moreover, they confirmed that DPC is a simple technique characterized by low-dissipation with cheap computation costs and is more effective and efficient than the consistent form of PS. This technique employs a dynamic pair-wise repulsive force and a variable coefficient of restitution for the collision term for dealing with different conditions of inter-particle penetrations. They showed that for the case of MPS, unlike the widely used Particle Shifting (PS) method, the DPC transport-velocity equation conserves the linear momentum of the system. Moreover, the DPC technique is more straightforward as it is free from the complexities of interface treatments.

This paper aims at developing the DPS technique for the weakly compressible SPH method to demonstrate its role in the accuracy and stability of the highly dynamic free-surface flow modeling. It also represents and evaluates a hybrid PS-DPC technique. We re-derive the general form of the DPC algorithm in the SPH framework for multiphysics problems (Section \ref{sec:DPC}). We append DPC to the standard form of the PS equation (as a hybrid technique) to resolve the particle clustering issue present on the free-surface and external particles while exempting the model from complex free-surface treatments (Section \ref{sec:cPS}). We implement these techniques in the GPU-accelerated subroutines of the open-source code DualSPHysics (see \cite{Crespo2015} and \cite{Dominguez2021}). We simulate four benchmark cases: (i) the oscillating droplet under a conservative force field, (ii) the two-dimensional (2D) water dam-break, (iii) the 2D water sloshing in a tank, and (iv) the three-dimensional (3D) water dam break against a rigid obstacle (Section \ref{sec:Results}). By qualitative and quantitative validations, we evaluate the convergence and consistency of SPH with the DPC technique compared to the existing and newly developed PS algorithms.

\section{SPH methodology}
\label{sec:SPH}
In SPH, the Lagrangian form of the Navier-Stokes equations represent the physical laws of the fluid flows \cite{Liu2003}. The numerical method discretizes the governing equations over the entire computational domain, $ \Omega $, into moving particles (or simply particles) categorized as the fluid particles, $ \Omega_f $, and the solid boundary particles, $ \Omega_b $ (i.e., $\Omega= \Omega_f \cup \Omega_b $). The approximation operator,  $\langle\centerdot\rangle$, forms the continuity and momentum equations, respectively, to:
\begin{equation}\label{eq:Cont}
	\frac{\mathrm{D}\rho_i}{\mathrm{D}{t}}=-{\rho}_i{\langle\nabla\cdotp\mathbf{v}\rangle}_i
\end{equation}
and
\begin{equation}\label{eq:Mom}
     {\rho_i}\frac{\mathrm{D}\mathbf{v}_i}{\mathrm{D}{t}}=-\langle\nabla{p}\rangle_i+\langle\nabla\cdotp\uptau\rangle_i+{\rho_i}{\mathbf{F}_i},
\end{equation}
in which the material time derivative, ${\mathrm{D}\mathbf({\centerdot})}/{\mathrm{D}{t}} $, updates the density, $ \rho_i $, and velocity, $ \mathbf{v}_i $, of the generic particle, $ i \in \Omega_f$. In the momentum equation (\ref{eq:Mom}), the pressure, the total shear stress tensor, and the body force per unit volume, are denoted as $ p_i $, $ \uptau_i $, and ${\rho_i}{\mathbf{F}_i}  $ respectively. The fluid particle carries the material and flow properties as its local position, $ \mathbf{r}_i $, moves by
\begin{equation}\label{eq:dis}
	\frac{\mathrm{D}\mathbf{r}_i}{\mathrm{D}{t}}={\mathbf{v}_i}.
\end{equation} 
Considering the fluid phase as a weakly compressible and barotropic fluid, the equation of state,
\begin{equation}\label{eq:EOS}
	p_{i}=B_0 \left(\left(\frac{\rho_i}{\rho_0}\right)^\gamma-1\right),
\end{equation}
explicitly updates $ p_i $ where the fluid bulk modulus is $ B_0={{c}^2_s\rho_0}/{\gamma} $. The reference density of the fluid and the polytropic index are denoted as $ \rho_0 $ and $ \gamma=7 $, respectively. With the numerical speed of sound, $ c_s $, being much greater than the maximum expected velocity, $ {\|\mathbf{v}\|}_{max} $, this model restricts the density variations, e.g., with $ c_s\geqslant10{{\|\mathbf{v}\|}_{max}} $, the Mach number, $ Ma \leqslant 0.1$, and the density variation would be limited to less than \%1. 

In the variationally consistent framework, the SPH formulation estimates the divergence of velocity, $ {\langle\nabla\cdotp\mathbf{v}\rangle}_i $, and the gradient of pressure, $ \langle\nabla{p}\rangle_i $, terms as follows \cite{Dominguez2021}:
 \begin{equation} 
	\left\{
		\begin{array}{l}
		\displaystyle
		\langle{\nabla\cdotp\mathbf{v}}\rangle_i={\sum_{j}\frac{m_{j}}{\rho_{j}} {\mathbf{v}_{ji}}\cdotp\nabla_i{W}_\mathit{ij}}\\[18pt]
		\displaystyle
		{\langle\nabla{p}\rangle_i}={\sum_{j}\frac{m_{j}}{\rho_{j}} \left({p}_{i}+{p}_{j}\right)\nabla_i{W}_\mathit{ij}}
     	\end{array}
	 \right.\label{eq:AppTerms}
\end{equation}
in which $W_\mathit{ij}= W(r_\mathit{ij}, kh) $ is the smoothing kernel function, $ m_i $ denotes the constant mass of the particle, and $ (\centerdot)_\mathit{ij}=(\centerdot)_i-(\centerdot)_j $. The neighbor particle, $ j \in \Omega $, is within the support domain of $ W $ with the influence radius of $ kh $, i.e., the inter-particle distance, ${r}_\mathit{ij}=\|{\mathbf{r}_{i}-\mathbf{r}_{j}}\|\leq{kh} $ (where $ k $ is a positive real number and the smoothing length, $ h=2dp $, $ dp $ being the initial particle distance). The gradient of kernel, $ \nabla_i W_\mathit{ij}=({\mathbf{r}_\mathit{ij}}/{r_\mathit{ij}})({\partial W_\mathit{ij}}/{\partial r_\mathit{ij}}) $ noting that $ {\partial W_\mathit{ij}}/{\partial r_\mathit{ij}} \leq0 $ \cite{Liu2003,Monaghan2005}. The shear force in (\ref{eq:Mom}), i.e., $ \langle\nabla\cdotp\uptau\rangle_i $, includes the momentum dissipation due to the laminar flow regime and the turbulence effects. Adopting the Laplacian and the variationally consistent approximation operators, respectively, for the laminar viscosity and the turbulence dissipation terms, gives \cite{Crespo2015, Dominguez2021}:
\begin{equation}\label{eq:LamSPS}
	{\langle\nabla\cdot\uptau\rangle_i}={\sum_{j}m_{j}\frac{4\rho_i\nu_0\mathbf{r}_\mathit{ij}\cdotp\nabla_i{W}_\mathit{ij}}{(\rho_i+\rho_j)(r^2_\mathit{ij}+\eta^2)}\mathbf{v}_\mathit{ij}}+{\sum_{j}\frac{m_{j}}{\rho_{j}} \left(\uptau^{*}_{i}+\uptau^{*}_{j}\right)\cdotp\nabla_i{W}_\mathit{ij}}.
\end{equation}
in which $ \nu_0 $ is the reference kinematic viscosity of the fluid and $ \eta=0.01h $ is a small value added to avoid the singularity issue. The large eddy simulation sub-particle scale model (SPS), using the Favre averaging for the weakly compressible model, determines the turbulence stress tensor, $ \uptau^{*}_i $, as follows \cite{Gotoh2001, Dalrymple&Rogers2006}:
\begin{equation}\label{eq:SPS}
	\uptau^{*}_i=2{\nu_t}_i\rho_i\left(\mathbf{S}_i-\frac{1}{3}{tr}(\mathbf{S}_i)\mathbf{I}\right)-\frac{2}{3}\rho_iC_I\Delta^2|\mathbf{S}_i|^2\mathbf{I}
\end{equation}
in which $ \mathbf{S}_i $ is the strain rate tensor, $ \mathbf{I} $ is the identity matrix, $ {\nu_t}_i=(C_s\Delta)^2|\mathbf{S}_i| $ is the eddy viscosity, and  $ C_I=0.00066 $. The Smagorinsky coefficient, $ C_s=0.12 $, and the filter width, $ \Delta $ (which is as a constant proportional to $ dp $) determine the mixing length-scale of the SPS turbulence model. The magnitude of the strain rate tensor, denoted as $ |\mathbf{S}_i| $, reads  $ \sqrt{2S^{\alpha\beta}_iS^{\alpha\beta}_i} $ (in Einstein notation where $ S^{\alpha\beta}_i $ is an element of $ \mathbf{S}_i $ considering $\alpha$ and $ \beta $ as the coordinate directions).

Appending an artificial density diffusion term to the right-hand side of the continuity equation (\ref{eq:Cont}) has been an effective and essential solution (see the $ \delta $-SPH method of Molteni and Colagrossi \cite{Molteni&Colagrossi2009} and the enhanced MPS method of Jandaghian and Shakibaeinia \cite{Jandaghian2020}) to eliminate high-frequency pressure noises of the weakly-compressible model\RvMJ{s}. In the weakly compressible SPH formulation, the general form of the density diffusion term, $ D_i $, is written as follows:
\begin{equation}\label{eq:Di}
	D_i=\delta h c_s \sum_{j\in \Omega_f}\frac{m_{j}}{\rho_{j}}\mathbb{\uppsi}_\mathit{ij}\cdotp\nabla_i{W}_{ij}
\end{equation}
where the preset non-dimensional coefficient, $  \delta $, adjusts the magnitude of this numerical correction. The standard form of the density diffusion term is given by \cite{Molteni&Colagrossi2009}:
\begin{equation}\label{eq:Molt}
	{\uppsi}_\mathit{ij}=2(\rho_j-\rho_i)\frac{\mathbf{r}_\mathit{ji}}{(r^2_\mathit{ij}+\eta^2)}.
\end{equation}
This term causes the density field close to the boundaries to diverge from the hydrostatic solution \cite{Antuono2010, Jandaghian2021_AWR}. To improve the pressure estimation at the vicinity of the solid boundary, Fourtakas et al. \cite{Fourtakas2019}, proposed a new form of (\ref{eq:Molt}) by replacing the total density with the dynamic density as: 
\begin{equation}\label{eq:Fourt}
		{\uppsi}_\mathit{ij}=2(\rho_\mathit{ji}-\rho^H_\mathit{ij})\frac{\mathbf{r}_\mathit{ji}}{(r^2_\mathit{ij}+\eta^2)}.
\end{equation}
The hydrostatic pressure difference, $p^H_\mathit{ij}=\rho_0\mathbf{g}\cdotp\mathbf{r}_\mathit{ji} $, gives the hydrostatic density difference, $ \rho^H_\mathit{ij} $, through the inverse form of the equation of state as \cite{Fourtakas2019}:
\begin{equation}\label{eq:rhoH}
	 \rho^H_\mathit{ij} =\rho_0\left(\sqrt[\gamma]{1+\frac{p^H_\mathit{ij}}{B_0}}-1\right),
\end{equation} 
with $ \mathbf{g} $ being the gravitational acceleration set to $ \{0,0,-g=-9.81 m.s^{-2}\}^T $.

To solve the governing equations (\ref{eq:Cont}-\ref{eq:EOS}), we adopt in this study the second-order symplectic scheme in which the explicit model dynamically updates the time steps of calculation, $ \Delta t $, based on the Courant–Friedrichs–Lewy (CFL) stability conditions (see \cite{Dominguez2021}). We employ the modified Dynamic Boundary Condition (mDBC) method implemented by English et al. \cite{English2021} for solid boundaries.

\section{Particle regularization techniques}\label{sec:PRT}
\subsection{Dynamic Particle Collision (DPC) technique for SPH}\label{sec:DPC}
\begin{figure}[H]
	\centering
	\includegraphics[width=\textwidth]{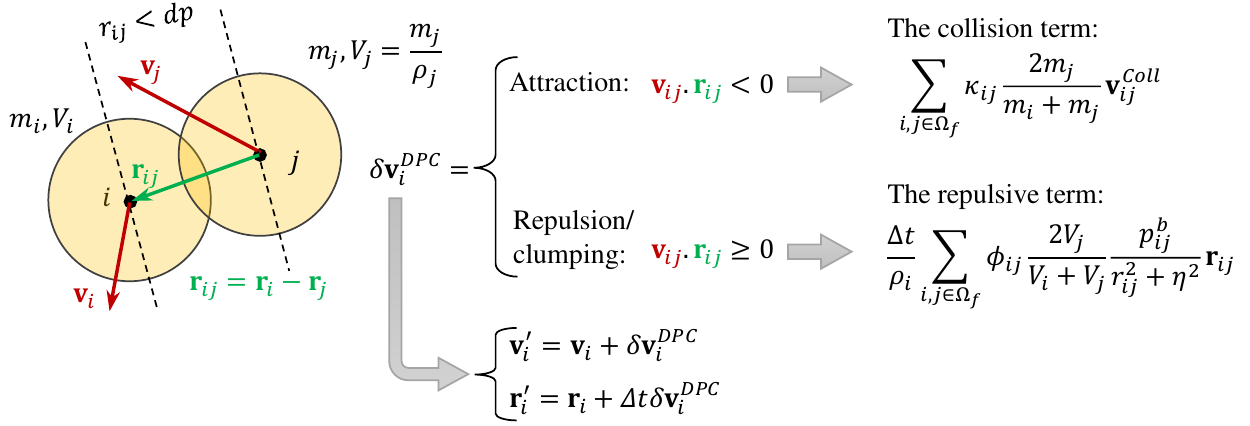}
	\caption{The general form of the DPC transport-velocity formulation, $ \delta \mathbf{v}^\mathit{DPC}_i $, (\ref{eq:DPC}) consists of the dynamic collision and repulsive terms for dealing with different states of inter-particle penetration where $ r_{\mathit{ij}}<dp $}
	\label{fig_DPCeq}
\end{figure}

The pair-wise particle collision method (originally implemented in MPS by Lee et al. \cite{Lee2011} and Shakibaeinia and Jin \cite{Shakibaeinia_cmame_2012}) regularizes the particle distribution based on the momentum transfer between a pair of colliding particles. Jandaghian et al. \cite{Jandaghian2021_JCP} proposed a new version of this technique, denoted as the Dynamic pair-wise Particle Collision (DPC) technique, by adopting dynamic form of the collision and repulsive terms to improve the pressure field. Here, we represent the DPC formulation in the framework of the weakly compressible SPH method. Considering mass, $ m_i $, and volume, $ V_i=m_i/\rho_i $, of particles, the general form of the DPC transport-velocity equation, $ \mathbf{v}^\mathit{DPC}_i $, reads: 
\begin{equation}\label{eq:DPC}
	\delta \mathbf{v}^\mathit{DPC}_i=\sum_{j\in\Omega_f}\kappa_\mathit{ij}\frac{2{m}_\mathit{j}}{{m}_\mathit{i}+{m}_\mathit{j}}\mathbf{v}^{\mathit{coll}}_\mathit{ij}+
	\frac{\Delta t}{{\rho}_i}\sum_{j\in\Omega_f}\phi_\mathit{ij}\frac{2{V}_\mathit{j}}{{V}_\mathit{i}+{V}_\mathit{j}}\frac{p^b_\mathit{ij}}{r^2_\mathit{ij}+\eta^2}\mathbf{r}_\mathit{ij},
\end{equation} 
in which
\begin{eqnarray}\label{VColl}
	(\mathbf{v}^\mathit{coll}_\mathit{ij}, \phi_\mathit{ij}) = \left\{
	\begin{array}{ll}
		(\frac{\mathbf{v}_\mathit{ij} \cdot\mathbf{r}_\mathit{ij}}{{r^2_\mathit{ij}+\eta^2}}\mathbf{r}_\mathit{ji},0), & \mbox{for } \mathbf{v}_\mathit{ij} \cdot\mathbf{r}_\mathit{ij}< 0 \\[8pt]
		\left(0,1\right)         & \mbox{Otherwise}
	\end{array} \right.,
\end{eqnarray}
 and in single fluid phase simulations $ m_i=m_j $. The DPC transport-velocity equation consists of the collision and repulsive terms (i.e., the first and second terms at its right-hand side, respectively). It deals with different states of inter-particle penetration and uses a variable coefficient of restitution, $ \kappa_\mathit{ij} $, and the dynamic background pressure, $ p^b_\mathit{ij} $ \cite{Jandaghian2021_JCP} (see Fig. \ref{fig_DPCeq}). When a pair of particles overlap and approach each other (i.e., when $ r_\mathit{ij}< {dp} $ and $ \mathbf{v}_\mathit{ij} \cdot\mathbf{r}_\mathit{ij}< 0 $) the collision term reduces their normal collision velocity, $  \mathbf{v}^\mathit{coll}_\mathit{ij}$, and the binary multiplier $ \phi_\mathit{ij}=0 $. Otherwise, when $ \mathbf{v}_\mathit{ij} \cdot\mathbf{r}_\mathit{ij}\geq 0 $ and still the inter-particle penetration occurs, by setting $ \phi_\mathit{ij}=1$ and $  \mathbf{v}^\mathit{coll}_\mathit{ij}=0$, the model activates the repulsive force (derived from the pressure gradient term without any smoothing procedure). The dynamic background pressure of the repulsive term is limited to the expected maximum and minimum pressure\RvMJ{s} of the test case (denoted as $ p_{max} $ and $ p_{min} $, respectively) and is written as \cite{Jandaghian2021_JCP}:
\begin{eqnarray}\label{eq:Pb}
	\left\{
	\begin{array}{l}
		p_\mathit{ij}^{b} = \tilde{p}_\mathit{ij} \chi_\mathit{ij} \\[8pt]
		\tilde{p}_\mathit{ij} = \max\left( \min\left( \lambda\left\vert p_{i} + p_{j}
		\right\vert,\lambda p_{\mathit{max}} \right), p_{\mathit{min}}
		\right)
	\end{array}
	\right.
\end{eqnarray} 
where $ \lambda $ is a non-dimensional adjusting parameter (set to $ 0.1$). DPC dynamically determines the variable coefficients, $ \chi_{ij} $ and $ \kappa_\mathit{ij} $, through:

\begin{equation}\label{kij}
	\chi_\mathit{ij} = \sqrt{\frac{w( r_\mathit{ij},dp )}{w( 0.5dp, dp)}}\mbox{ and } \kappa_\mathit{ij}=\left\{
	\begin{array}{ll}
		\chi_\mathit{ij}&0.5 \leq r_\mathit{ij}/{dp} < 1\\[5pt]
		1 &      r_\mathit{ij}/{dp} <0.5
	\end{array} \right.,
\end{equation}
respectively. These coefficients control the strength of the repulsive and collision terms as functions of the inter-particle distance via the non-dimensional part of the Wendland kernel given as:
\begin{eqnarray}\label{Wij}
	w(r_\mathit{ij}, {dp})= (1-\frac{r_\mathit{ij}}{dp})^4(4{\frac{r_\mathit{ij}}{dp}}+1), 0\leq r_\mathit{ij} < dp
\end{eqnarray} 
(where for $ r_\mathit{ij}\geq{dp}$, $ w(r_\mathit{ij}, {dp})=0 $). Eventually, DPC updates the velocity and position of the fluid particles by
 \begin{equation} 
	\left\{
	\begin{array}{l}
		\displaystyle
			\mathbf{v}^\prime_i=\mathbf{v}_i+\delta \mathbf{v}^\mathit{DPC}_i\\[6pt]
		\displaystyle
		\mathbf{r}^\prime_i=\mathbf{r}_i+\Delta{t} \delta \mathbf{v}^\mathit{DPC}_i
	\end{array}.
	\right.\label{eq:DPC2}
\end{equation}
We should highlight that \RvMJ{the DPC formulations (\ref{eq:DPC}-\ref{kij}) keep the velocity corrections small by implementing pair-wise particle interactions and the dynamic coefficients. Therefore, DPC avoids excessive manipulation of the mechanical properties of the flow (including the mass and volume of particles) without the need for the additional diffusion terms due to the non-Lagrangian velocity changes (derived based on the Leibniz–Reynolds transport theorem in the consistent PS formulations (e.g., \cite{Jandaghian2021_JCP, Sun2019}) and the Arbitrary Lagrangian Eulerian (ALE) schemes coupled with the PS equation (e.g., \cite{Oger2016, Antuono2020})).} Unlike standard PS methods that shift fluid particles to the area with less concentration (e.g., \cite{Lind2012, Sun2017_dplus}), DPC conserves the linear momentum of the two colliding particles by being an anti-symmetric formulation \RvMJ{and considering a constant mass for the particles \cite{Jandaghian2021_JCP}}. Furthermore, DPC implements the regularization process of the fluid particles without any boundary treatments eliminating the complexities of the free surface and normal vectors detection. \RvMJ{In  section \ref{sec:Results}, simulating benchmark cases, we evaluate the evolution of system's global energy to confirm the negligible effects of the DPC transport-velocity equation on the overall flow properties.} The implementation of DPC in DualSPHysics is represented in \ref{sec:appendixDPC}.

\subsection{Particle shifting algorithm coupled with DPC: \textit{cPS}}\label{sec:cPS}
The standard Particle Shifting (PS) technique, implemented in DualSPHysics, regularizes the distribution of fluid particles ($ i \in \Omega_f $) using the Fickian-based formulation (proposed by Lind et al.\cite{Lind2012} and Skillen et al. \cite{Skillen2013}):
\begin{equation}\label{eq:dr}
	\delta\mathbf{r}^\mathit{PS}_i=-{D^F_i}\langle{\nabla{C}}\rangle_i
\end{equation}
where the particle shifting vector, $ \delta\mathbf{r}^\mathit{PS}_i $, has the same direction as the gradient of the local particle concentration, $ \langle{\nabla{C}}\rangle_i $, estimated as:
\begin{equation}\label{eq:gCi}
	\langle{\nabla{C}}\rangle_i={\sum_{j\in\Omega}\frac{m_{j}}{\rho_{j}}\nabla_i{W}_\mathit{ij}}.
\end{equation}
A variable Fickian diffusion coefficient, $ {D^F_i}=Ah\Delta{t}\|\mathbf{v}_i\| $, controls the intensity of shifting based on the Von Neumann stability analysis and the CFL condition \cite{Skillen2013}. The non-dimensional constant, $ A $, is suggested to be set between 1-6 \cite{Skillen2013}. One should note that $ D^F_i $ is a function of the variable time step and the local velocity magnitude (unlike the \RvMJ{other} PS formulations that employ constant $ D^F $ using the maximum expected velocity of the test case, e.g., in \cite{Jandaghian2020, Sun2017_dplus}). Also, the model limits the magnitude of $ \delta \mathbf{r}^\mathit{PS}_i $ in each coordinate directions to $ 0.1dp $. Accordingly, this PS algorithm, through the variable Fickian diffusion coefficient and limiting the magnitude of shifting, aims at avoiding excessive movement of particles \cite{Dominguez2021}.

Inside the free-surface region, the kernel truncation causes (\ref{eq:dr}) to introduce false diffusion of particles toward and beyond the interface \cite{Lind2012, Sun2017_dplus}. Thus, special treatments of (\ref{eq:dr}) become necessary to control particle shifting at the free-surface and its vicinity. This includes detecting the free-surface particles and modifying/canceling the shifting normal to the interface. Accuracy of the particle shifting corrections greatly depends on the efficiency of the particle detection algorithm and the estimated normal vectors (especially in violent free-surface flows) \cite{Sun2017_dplus}. With inaccurate estimation of normal vectors and canceling the component of PS normal to the interface, the highly dynamic flows would be susceptible to particle clustering at the interface \cite{Jandaghian2020}. 

The standard PS method, for single-phase flow simulations, simply adjusts the magnitude of shifting by multiplying $ \delta\mathbf{r}^\mathit{PS}_i $ by a correction coefficient, $ {A_C}_i $. It also neglects the shifting of the particles with extreme kernel truncation (i.e., $ \delta\mathbf{r}^\mathit{PS}_i=0 $ for $ {A_C}_i <0$). The correction coefficient is defined as $ {A_C}_i=(\langle{\nabla\cdotp\mathbf{r}}\rangle_i-A_\mathit{FST})/(dim.-A_\mathit{FST}) $ in which
\begin{equation}\label{eq:divR}
	\langle{\nabla\cdotp\mathbf{r}}\rangle_i={\sum_{j\in \Omega}\frac{m_{j}}{\rho_{j}}{\mathbf{r}_{ji}}\cdotp\nabla_i{W}_\mathit{ij}},
\end{equation}
and the free-surface threshold coefficient, $ A_\mathit{FST}$, is a preset value less that the domain dimension, $dim.= 2 $ or $ 3 $ for the two- and three-dimensional simulations, respectively \cite{Lee2008, Mokos2017}. \RvMJ{We refer to the equations (\ref{eq:dr}-\ref{eq:divR}) as the standard PS formulation (which exists in DualSPHysics)}.

\begin{figure}[H]
 	\centering
 	\includegraphics[width=\textwidth]{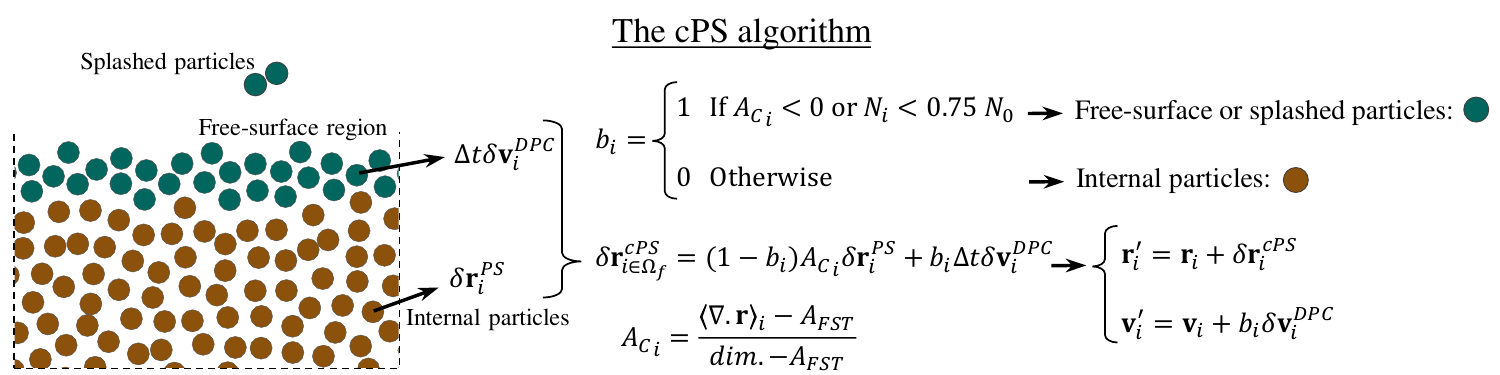}
 	\caption{The coupled Particle Shifting formulation (\ref{eq:cPS}). $ b_i $, $ \delta\mathbf{r}^\mathit{PS}_i $, and $ \delta\mathbf{v}^\mathit{DPC}_i $ are given through equations (\ref{eq:bi}), (\ref{eq:dr}), and (\ref{eq:DPC}), respectively.}
 	\label{fig_cPS}
\end{figure}

To improve particle distribution in the free-surface region and ensure numerical instability, Jandaghian et al. \cite{Jandaghian2021_JCP, Jandaghian2020} coupled the PS formulation with the particle collision technique within the context of MPS method. Here, similar to the work of Jandaghian et al. \cite{Jandaghian2020}, we couple the standard PS equation with the DPC technique, represented in section (\ref{sec:DPC}). In this hybrid method, hereinafter denoted as cPS, the PS equation (\ref{eq:dr}) is directly applied to the internal particles, while the DPC transport-velocity term (\ref{eq:DPC}) regularizes the distribution of particles detected inside the free-surface region or as splashed particles. We first classify the fluid particles into two categories identified by a binary multiplier, $ b_i $, given as:
 \begin{eqnarray}\label{eq:bi}
	b_{i} = \left\{
	\begin{array}{ll}
		1 & \mbox{If } {A_C}_i<0 \mbox{ or } N_i< 0.75N_0 \\[8pt]
		0 & \mbox{Otherwise}
	\end{array} \right..
\end{eqnarray}
in which $ N_i $, and $ N_0 $ are the number of neighbor particles and the expected number of neighbor particles in an isotropic particle distribution, respectively. If $ b_i=0 $ the model considers the fluid particle as an internal particle; if $ b_i=1 $ the particle belongs to the free-surface region or the splashed (i.e., external) particles' category due to the extreme kernel truncation and insufficient number of neighbor particles (see Fig. \ref{fig_cPS}). Considering $ k=2 $ in the smoothing functions, we estimate $ N_0=52 $ and $ 260 $ for the 2D and 3D simulations, respectively. Employing the binary multiplier (\ref{eq:bi}) and the DPC term (\ref{eq:DPC}), we implement the cPS equation as: 
\begin{equation} 
	\delta\mathbf{r}^\mathit{cPS}_{i\in \Omega_f}=(1-b_i){A_C}_i\delta\mathbf{r}^\mathit{PS}_i+b_i{\Delta{t}}\delta{\mathbf{v}^\mathit{DPC}_i}
	\label{eq:cPS}
\end{equation}
which finally updates the position and velocity of the fluid particles through 
\begin{equation} 
	\left\{
	\begin{array}{l}
		\displaystyle
		\mathbf{r}^\prime_i=\mathbf{r}_i+\delta \mathbf{r}^\mathit{cPS}_i\\[6pt]
		\displaystyle
		\mathbf{v}^\prime_i=\mathbf{v}_i+b_i\delta\mathbf{v}^\mathit{DPC}_i
	\end{array}
	\right.\label{eq:cPS2}.
\end{equation} 
We implement cPS only in the correction stage of the time integration scheme (compatible with the DPC implementation shown in Fig. \ref{fig_DPC}). This coupling of PS with DPC exempts the model from complex free-surface treatments which would involve time-consuming operations (required for estimating the normal vectors and the renormalization tensor), noting that DPC improves the particle clustering issue in the free-surface and splashed regions (i.e., where $ b_i=1 $). Fig. \ref{fig_cPS} summarizes the cPS algorithm. It should be highlighted that the proposed cPS method still lacks the consistent particle shifting algorithms developed by implementing additional diffusion/cohesion terms to encounter the volume-non-conservation issue of the standard PS models (e.g., see \cite{Hong-Guan2021, Jandaghian2021_JCP, Sun2019}).
\section{Numerical simulations}\label{sec:Results}
We simulate challenging benchmark cases to investigate the stability and accuracy of the weakly compressible SPH method supplied with the proposed particle regularization techniques. We choose numerical cases in which violent free-surface flows and the potential forces are dominant. Through a comprehensive evaluations of the numerical results, we discuss the effectiveness and efficiency of the proposed models for capturing the impact events and long-term modeling of highly dynamic flows. For all the simulations, we adopt the fifth-order Wendland kernel and set $ k=2  $. In the symplectic time integration scheme the CFL coefficient is set to 0.2. In the PS formulation, we use the recommended values of the shifting coefficients, i.e., $ A =2 $ and $ A_\mathit{FST} = 1.5 $ and $ 2.75 $ for the 2D and 3D problems, respectively \cite{Dominguez2021}. To quantitatively study the effectiveness of the particle regularization techniques, we calculate and plot the spatial particle disorder, denoted as $ \lambda_i $ (which its formulations are proposed by Antuono et al. \cite{Antuono2014}). By averaging $ \lambda_i $ over the internal fluid particle (i.e., for $ b_i=0 $ given by (\ref{eq:bi})) the global value of the spatial particle disorder, $ \Lambda $, is estimated \cite{Antuono2014}. Moreover, the evolution of global energy is identified as $ \Delta E=E^t-E^0 $ (where $ E^t $ and $E^0$ refer to the global energy at $ t $ and the initial condition); the global potential, kinetic, and mechanical energies, are denoted as $ E_p $, $E_k$, and $ E_m $, respectively (see \cite{Jandaghian2021_JCP} for their general formulation). Videos of the simulations are provided in the supplementary material \ref{sec:appendixSupp}.

Herein, we denote the weakly compressible SPH model (Section (\ref{sec:SPH})) coupled either with the proposed DPC formulation (Section (\ref{sec:DPC})), the cPS algorithm (Section (\ref{sec:cPS})), or the standard PS method as \textit{DPC}, \textit{cPS}, or \textit{PS}, respectively. 
 
\begin{figure}[H]
	\centering
	\includegraphics[width=\textwidth]{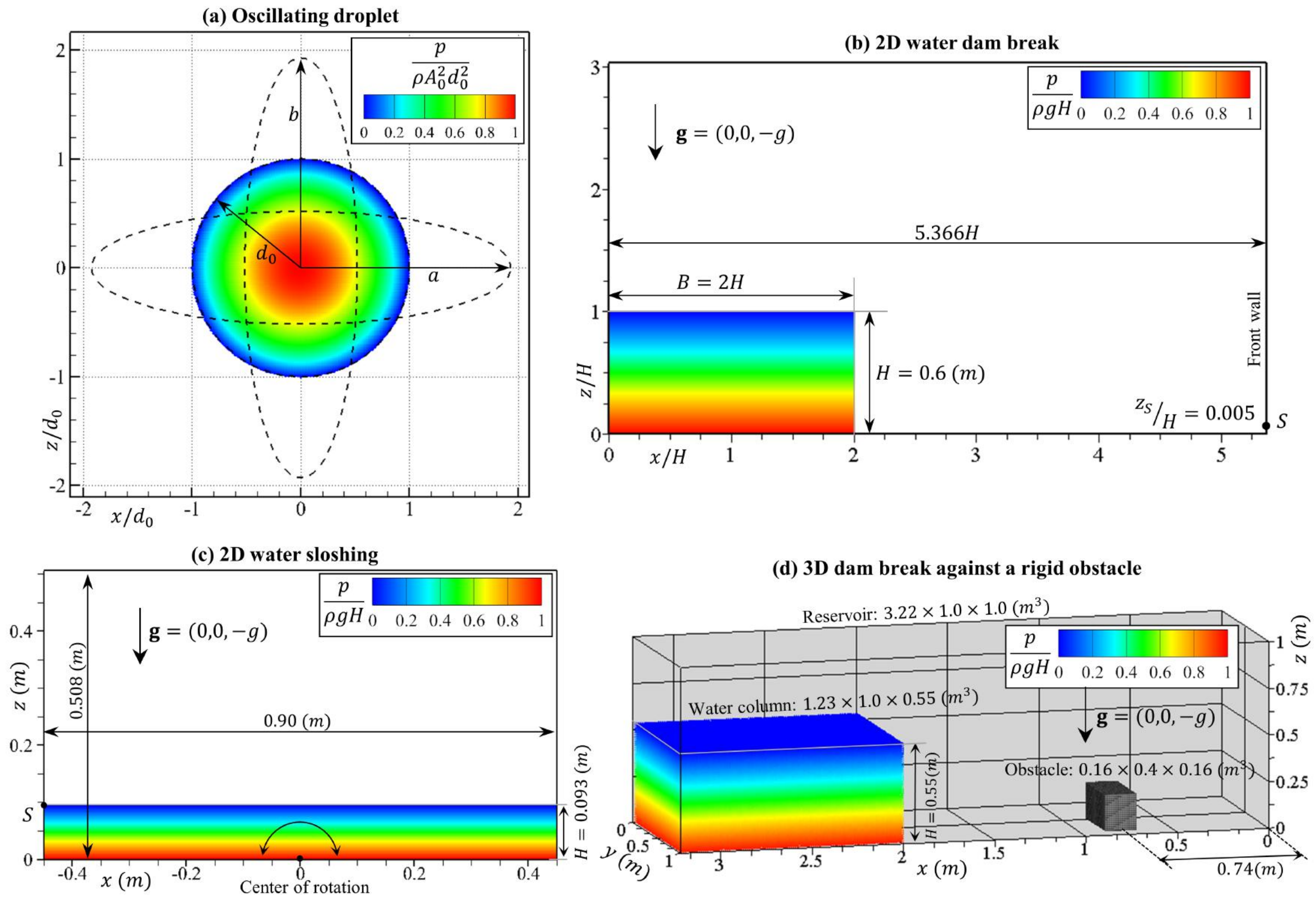}
	\caption{Initial state and pressure field of the numerical test cases: (a) the oscillating droplet, (b) 2D water dam break, (c) 2D water sloshing, and (d) 3D dam break against an obstacle}
	\label{fig_Initial}
\end{figure}

We represent flow characteristics and numerical properties of the test cases as follows:
\begin{enumerate}
	
\item {\textit{Oscillating droplet under a conservative force field:}} This numerical test case is a periodic free-surface flow. By long-term simulations of this benchmark case, we can directly investigate the conservative properties of the particle method considering that no solid boundary exists and the fluid is inviscid \cite{Sun2017_dplus, Antuono2015}. A conservative central body force, $ \mathbf{F}_i=-A_0^2 \mathbf{r}_i $, periodically stretches an initially circular droplet, with the radius of $ d_0 $, along the \textit{x} and \textit{z}-directions (Fig. \ref{fig_Initial}-a). The flow frequency, $ 1/T\simeq A_0/4.827 $, and the constant parameter $ A_0=1.5 $ ($ 1/s $) control the time evolution of this free-surface flow. We distribute the particles on the Cartesian lattice where $d_0=0.5$ ($m$) and assign the initial condition based on the theoretical solution by Monaghan and Rafiee \cite{Monaghan2013} (i.e., $p_i= \rho_0 A_0^2(d_0^2-\|{\mathbf{r}_i}\|^2) $ and $ \mathbf{v}_i=\{A_0,0,-A_0\}{\mathbf{r}_i}$ at $ t=0 $). Moreover, we set $ c_s=15A_0 d_0 $ and $ \rho_0=1000$ ($ kg/m^3  $) and identify the spatial resolution of the test case by $ R=d_0/dp $. To avoid high-frequency pressure noises, we only activate the density diffusion term of Molteni and Colagrossi \cite{Molteni&Colagrossi2009} (\ref{eq:Molt}) with $ \delta=0.1 $ (without the artificial viscosity term as $ \nu_0=0 $). We assign the initial maximum and minimum pressures at the center and the free-surface of the droplet to $ p_{max}=2\rho_0 A_0^2d_0^2 $ and $ p_{min}=(\rho_0 A_0^2(d_0^2-(d_0-0.5dp)^2) $ in the DPC repulsive term (\ref{eq:Pb}), respectively. 

\item {\textit{2D dam break:}} This problem has become a popular benchmark case for showing the robustness of the enhanced particle methods capable of simulating highly dynamic free-surface flows (e.g., \cite{Jandaghian2021_JCP, Molteni&Colagrossi2009,Lind2012, Hong-Guan2021, Liang-Yee2021}). It involves complex fluid-fluid and fluid-solid impacts that challenge the standard SPH methods for dealing with their associated numerical issues. Here, we simulate the experimental case by Lobovský et al. \cite{Lobovsky2014} in which the height and width of the water column are $ H= $ 0.6 ($ m $) and $ B=2H $, respectively, and the length of channel is $ 5.366H $ (in the $ x $-direction) (Fig. \ref{fig_Initial}-b). The initial hydrostatic pressure is assigned to the particles and the sound speed is set to $ c_0=10 \sqrt{gH} $ where the gravitational acceleration is $ g=9.81$ ($ m/s^2 $) in the negative $ z $-direction. \RvMJ{Unless specified}, the model implements the laminar viscosity and the SPS equations (\ref{eq:LamSPS}) where $ \nu_0=10^{-6}$ ($ m^2/s$) and $ \rho_0=1000$ ($ kg/m^3  $) for water. We activate the density diffusion model of Fourtakas et al. \cite{Fourtakas2019} (\ref{eq:Fourt}) and set $ \delta= 0.1 $.  The non-dimensional time and the spatial resolution of the model are identified by, $ T=t \sqrt(g/H) $ and $ R=H/dp $, respectively. For simulating the solid walls, we adopt the modified Dynamic Boundary Condition (mDBC) introduced by \cite{English2021}. Similar to \cite{Jandaghian2021_JCP}, we assign $ p_{max}=2.3\rho gH $ and $ p_{min}=\rho g dp $ in the DPC formulation (\ref{eq:Pb}). The impact load on the front wall, the probe location $ S $, is locally averaged to be compared with the experimental measurements.

\item {\textit{Water sloshing:}} Accurate estimation of the sloshing loads are essential for optimum design of fluid storage tanks and vessels \cite{Iglesias2011}. The movement of the solid boundary with the presence of the gravitational acceleration, $ \mathbf{g} $, forms plunging waves and lateral water impacts. Here, we simulate the water sloshing in a rectangular reservoir identical to the experimental case of Souto-Iglesias \cite{Iglesias2011}. In this test case, the tank rotates periodically under a sinusoidal excitation with the rotation center at the middle point of the bottom side and the frequency of $ 1/T $ (Fig. \ref{fig_Initial}-c). We simulate this test case as a 2D and single phase problem. The tank length and height, $ L_b $ and $ H_b $, are 0.90 and 0.508 ($ m $), respectively. Water fills the width of the reservoir with the initial height, $ H =0.093$ ($ m $). The model assigns the initial hydrostatic pressure to the particles and sets $ c_0=30 \sqrt{gH} $. The density and kinematic viscosity of the water are set to $ \rho=1000 $ ($ kg/m^3  $) and $ \nu_0=10^{-6}$ ($ m^2/s$), respectively. The SPS model calculates the shear forces in the numerical model. The density diffusion term of Fourtakas et al. \cite{Fourtakas2019} (\ref{eq:Fourt}) is implemented with $ \delta= 0.1 $. We adopt a single spatial resolution for all the simulations where $ dp=0.002 $ ($ m $). The mDBC model updates the pressure of the solid boundary particles. In (\ref{eq:Pb}), we set $ p_{max}=15$ ($ kPa $) and $ p_{min}=\rho g dp $ based on the expected maximum pressure from the experimental data and the minimum hydrostatistic pressure on the free-surface. We extract the local averaged pressure at the location of the pressure sensor, $ S $, (identified on Fig. \ref{fig_Initial}-c) to validate the lateral impact with those from the experiment.

\item {\textit{3D dam break against a rigid obstacle:}} In this test case, the water column collapses under the gravitational force, $ \mathbf{g} $, on the horizontal bed and impacts a rigid cuboid obstacle. Plunging jets form and fluid-solid interactions occur as water flows over the obstacle and impacts the front-wall of the reservoir. The 3D configuration of this free-surface violent flow challenges the developed particle method in capturing its highly dynamic deformations and impact events (e.g., \cite{Jandaghian2021_JCP,  English2021,Mokos2017, Rezavand2020}). Fig. \ref{fig_Initial}-d shows the initial hydrostatic pressure and the geometrical properties of the problem. The reservoir, water column and obstacle dimensions are chosen according to the experimental setup of Kleefsman et al. \cite{Kleefsman2005} employed for validating the flow evolution and impact loads. The vertical lines probed for the fluid heights are at \textit{H1}:($ x=0.992 $, $ y=0.5 $) and \textit{H2}:($ x=2.638 $, $ y=0.5 $) and the local pressure is extracted on the front-vertical face of the obstacle at \textit{P1}:($ x=0.8245 $, $ y=0.471 $, $ z=0.021 $) and \textit{P2}: ($ x=0.8245 $, $ y=0.471 $, $ z=0.101 $) points (units in meter). The initial height of the water column is $ H =0.55$ ($ m $) in the z-direction. The non-dimensional time is denoted as, $ T=t\sqrt(g/H) $ and the sound speed is set to $ c_0=20 \sqrt{gH} $. Similar to the previous test cases, we implement the Laminar+SPS model (\ref{eq:LamSPS}) (with $ \rho=1000 $ ($ kg/m^3  $) and $ \nu_0=10^{-6}$ ($ m^2/s$)), the diffusion term of Fourtakas et al. \cite{Fourtakas2019} (\ref{eq:Fourt}) (with $ \delta= 0.1 $), and the mDBC solid boundary model. In the dynamic background pressure (\ref{eq:Pb}), we adopt $ p_{max}=2.3\rho gH$ and $ p_{min}=\rho g dp $. Simulating this 3D problem with various spatial resolutions ($ R=H/dp= 27.5$, $ 55 $, and $ 110 $), we specifically investigate the efficiency and computational performance of the new numerical implementations in DualSPHysics.

\end{enumerate}

\subsection{Oscillating droplet under a conservative force field}
\label{sec:OD}
\begin{figure}[H]
	\centering
	\includegraphics[width=\textwidth]{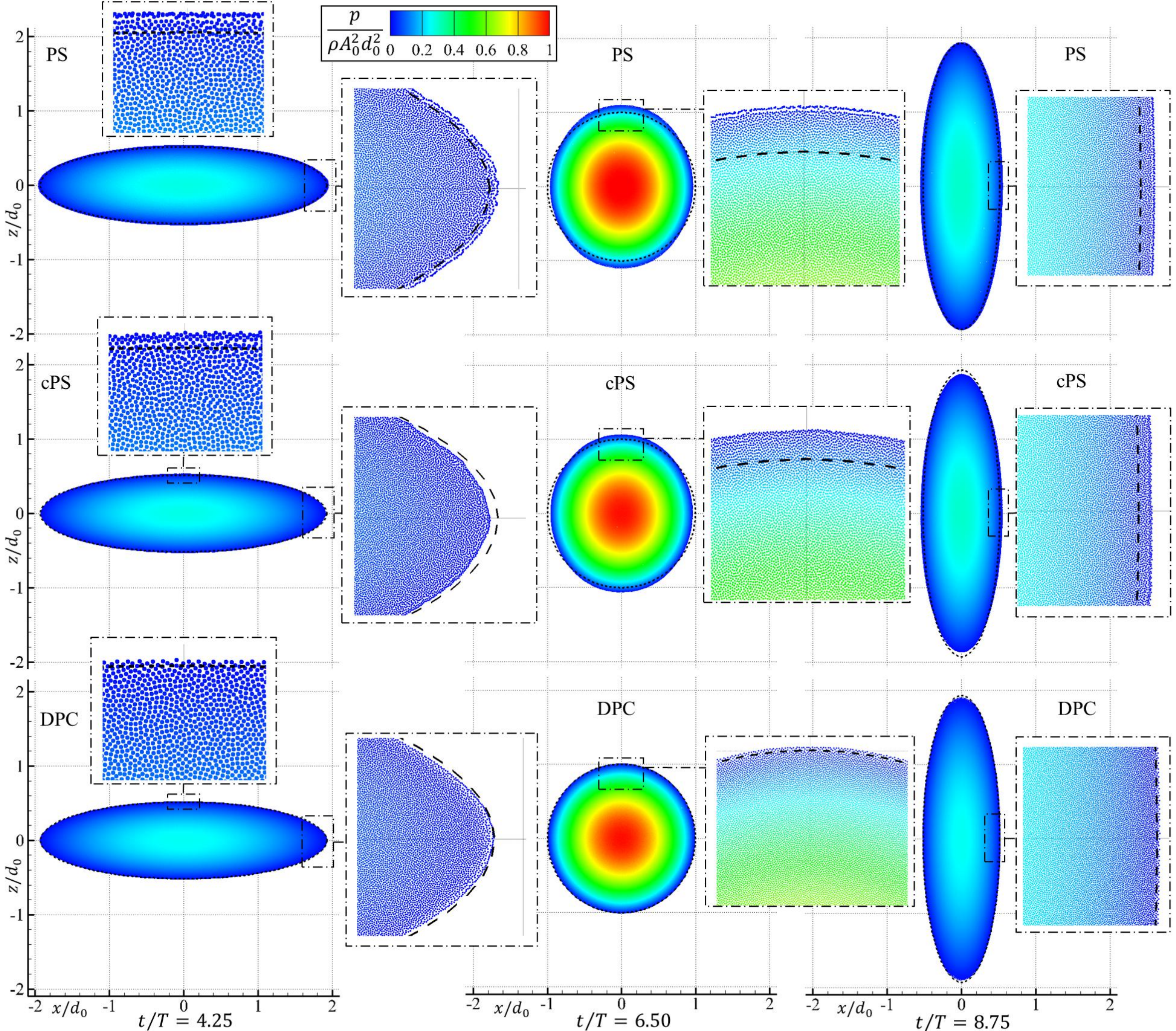}
	\caption{Oscillating droplet: non-dimensional pressure fields by the SPH with either PS, cPS, or DPC techniques from top to bottom rows, respectively \RvMJ{($ R=d_0/\mathit{dp}=200 $)}. The free-surface of the droplet is compared with the theoretical solution (the black dashed lines) at $ t/T=4.25, 6.50$, and $8.75 $.}
	\label{fig_ODPr}
\end{figure}

We validate the numerical simulations of this periodic free-surface flow versus its theoretical solution (represented in \cite{Monaghan2013}). Fig. \ref{fig_ODPr} illustrates the pressure fields and flow evolutions simulated with the PS, cPS, and DPC techniques, \RvMJ{where the spatial resolution, $ R={d_0}/\mathit{dp}=200 $}. Thanks to the density diffusion term, the pressure field is smooth over the fluid domain. However, the particle clustering issue at the free-surface region is notable with PS (as PS ignores the fluid particles with $ {A_C}_i<0 $). On the other hand, cPS improves particle distribution at the free-surface and its vicinity by applying DPC between the detected free-surface particles. Comparing the flow evolution with the analytical free-surface (indicated with the black dashed lines) manifests the unphysical volume expansion due to the inconsistent implementation of the particle shifting (particularly at $ t/T=$ 6.50 and 8.75); however, with cPS the volume expansion slightly reduces. The DPC method not only represents uniform particle distribution at the free-surface region ensuring the numerical stability, but also predicts accurate evolution of the oscillating droplet. 

\begin{figure}[H]
	\centering
	\includegraphics[width=4.0in]{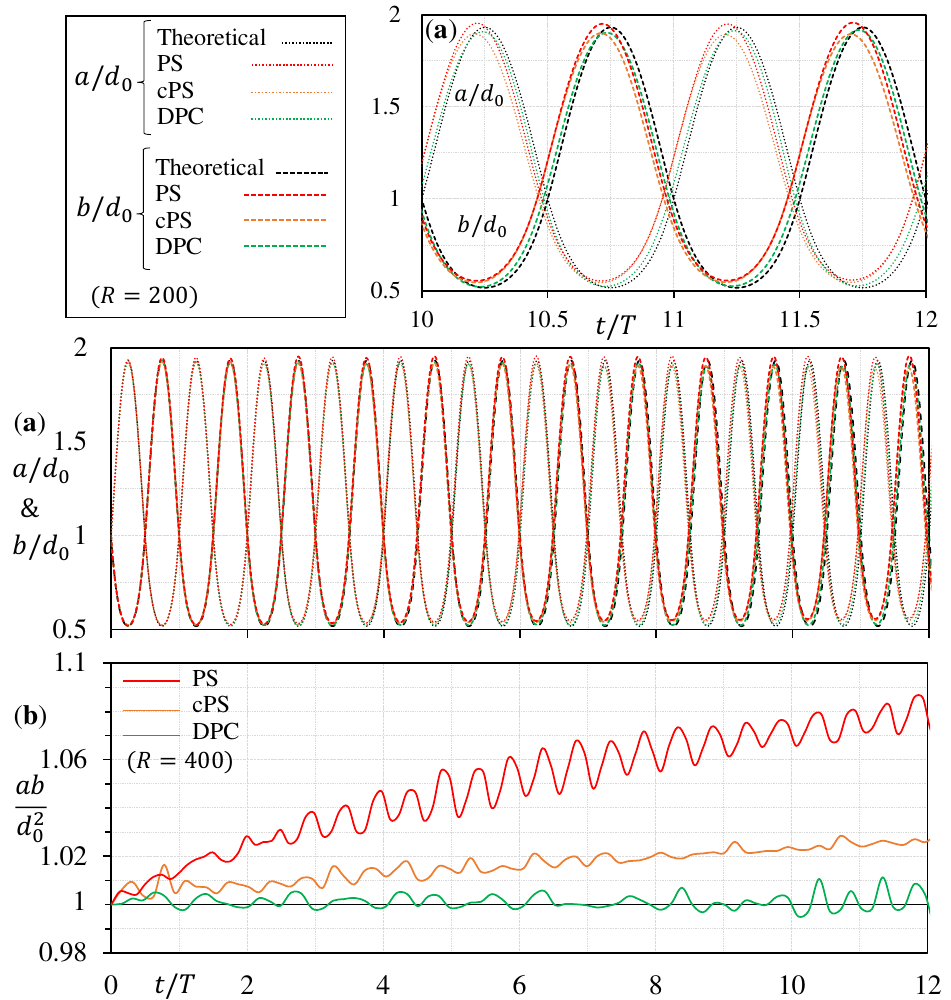}
	\caption{Oscillating droplet: the time evolutions of \textit{a} and \textit{b} by the SPH + PS, cPS or DPC techniques compared with the theoretical solution in (a). Graph (b) illustrates the divergence of model with the PS method due to the unphysical volume expansion}
	\label{fig_ODab}
\end{figure}

We plot the time evolution of semi-axes, $ a $ and $ b $, and their theoretical profiles to quantify the accuracy of the simulated flow. Fig. \ref{fig_ODab}-a shows that the shifting formulation leads to over overestimation of  $ a/d_0 $ and $ b/d_0 $, while the model with DPC gives more accurate evolution of the semi-axes. Theoretically, we expect the numerical model to conserve the total volume satisfying $ ab/d_0^2=1 $ condition. Fig. \ref{fig_ODab}-b shows that DPC successfully conserves the total volume with less than $ \sim $1 \% error. The naive implementation of PS diverges this value (i.e., the volume expands by $ \sim $8 \% after 12 cycles); nevertheless, the free-surface stability achieved by cPS reduces this error to less than 3 \%. 

\begin{figure}[H]
	\centering
	\includegraphics[width=\textwidth]{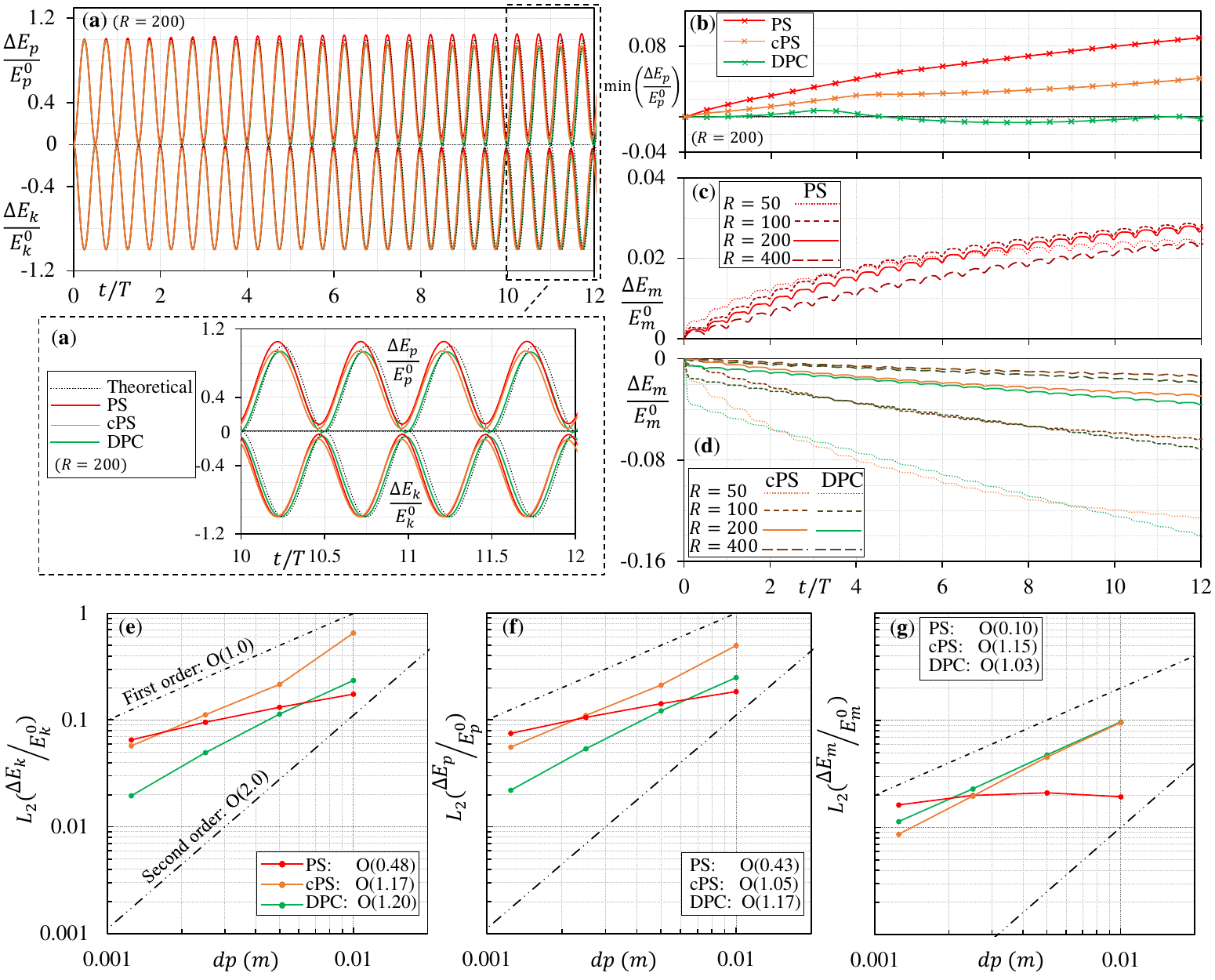}
	\caption{Oscillating droplet: graphs (a-d) represent the time evolution of the total potential ($ E_p $), kinetic ($ E_k $), and, mechanical ($ E_m $) energies comparing the numerical results with the theoretical solution (the black dotted lines). Graphs (e-g) give $ L_2  $ errors and the convergence order of the numerical results.}
	\label{fig_ODEn}
\end{figure}

In Fig. \ref{fig_ODEn}, we evaluate the time evolution of the global energies, $ \Delta E $. Fig. \ref{fig_ODEn}-a shows that DPC predicts more accurate evolution of the potential energy in comparison to PS that increases fluid volume over the simulation time. We observe that the minimum of potential energy evolution in each cycle, min$(\Delta E_p/E_p^0) $, (which should ideally be zero) diverges with PS by more than 8 \% error after 12 cycles, but cPS reduces the error to only $ \sim $4 \% (Fig. \ref{fig_ODEn}-b). On the other hand, DPC shows convergence behaviors as it keeps min$(\Delta E_p/E_p^0) $ to less than 1 \%. The global mechanical energy theoretically should remain identical to its initial value (i.e., ideally $ \Delta E_m=0 $). However, the numerical dissipation of the weakly compressible SPH algorithm (especially with implementing the density diffusion term) reduces the mechanical energy over the simulation (i.e., numerically $ \Delta E_m<0 $). Fig. \ref{fig_ODEn}-c \& d plot the mechanical energy evolution of the models with different spatial resolutions ($ R= $ 50, 100, 200, and 400). Fig. \ref{fig_ODEn}-d confirms that with DPC and cPS the mechanical energy converges toward its theoretical value as the spatial resolution increases. In contrast, PS results in positive values of mechanical energy evolution, i.e., $ \Delta E_m>0 $ (by inserting unphysival potential energy into the system). Also, PS does not establish a clear convergence behavior by increasing $ R $ (Fig. \ref{fig_ODEn}-c). The normalized root mean square error (i.e., $ L_2 $) of the kinetic, potential, and mechanical energies are plotted in Fig. \ref{fig_ODEn}, graphs (e), (f), and (g), respectively. In comparison to PS and cPS, the numerical error of the potential and kinetic energies significantly reducs by DPC. Yet, DPC and cPS result in almost identical order of convergence (1.0-1.2). Moreover, the PS affecting the global energies of the flow affects the convergence order reducing by one order of magnitude to 0.1-0.5.

\begin{figure}[H]
	\centering
	\includegraphics[width=\textwidth]{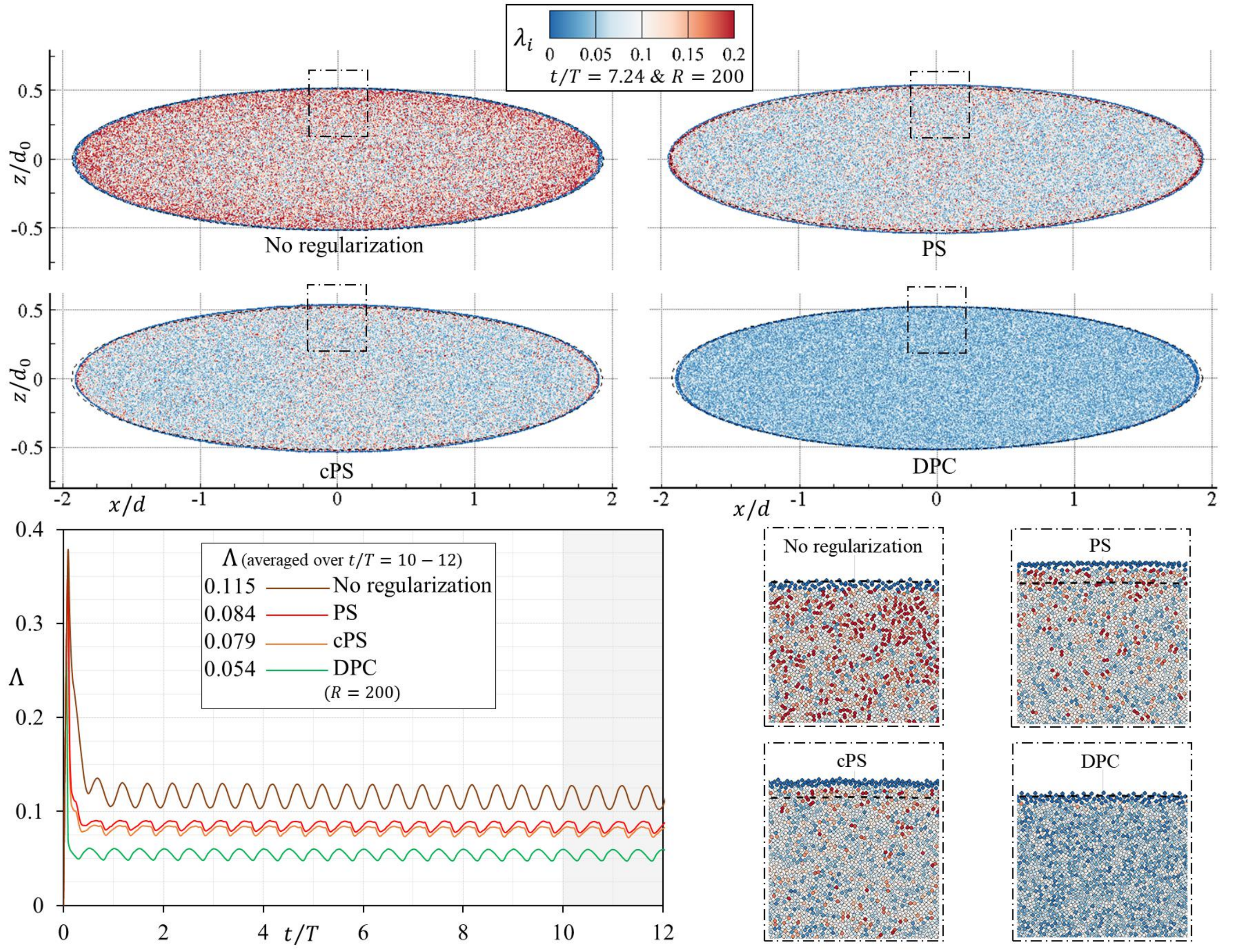}
	\caption{Oscillating droplet: the spatial particle disorder, $ \lambda_i $ for the models with no particle regularization, PS, cPS, and DPC. The time evolution of the global particle disorder, $ \Lambda $, is plotted in the graph for the different test cases. The spatial particle disorder formulation is implemented based on the work of Antuono et al. \cite{Antuono2014}.}
	\label{fig_Lambda}
\end{figure}

Here, we evaluate the local spatial particle disorder, $ \lambda_i $, and its global value $ \Lambda $ (represented by Antuono et al. \cite{Antuono2014}) comparing the effectiveness of the particle regularization techniques. The results show that the shifting formulation implemented in PS and cPS reduces $ \Lambda $ (averaged over $ t/T=10-12 $) from 0.115, obtained by the case with no particle regularization technique, to 0.084 and 0.079, respectively (Fig. \ref{fig_Lambda}). The improvement of particle distribution at the free-surface region with cPS is well illustrated in the zoom-in snapshots. It should be highlighted that DPC successfully represents more regular particle distribution ($ \Lambda=0.054 $) compared to the shifting technique.


\subsection{2D water dam break}
\label{sec:DB}
\begin{figure}[H]
	\centering
	\includegraphics[width=\textwidth]{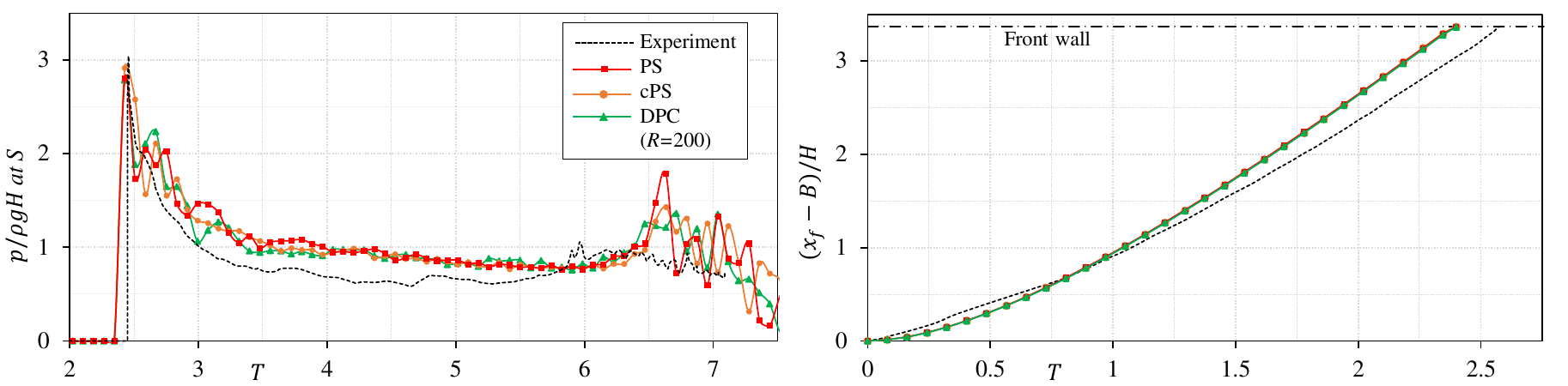}
	\caption{Dam break: the local averaged pressure on the front wall, at $ S $, (left) and the wave propagation on the horizontal bed, \RvMJ{$ {x_f} $}, (right) with \RvMJ{$ R=H/\mathit{dp}=200 $}. \RvMJ{Numerical results are extracted every 0.02 seconds.}}
	\label{fig_DBExp}
\end{figure}

We validate the numerical simulation of the 2D dam break versus the experimental results of Lobovský \cite{Lobovsky2014} and evaluate the role of the particle regularization techniques. Fig. \ref{fig_DBExp} shows that DPC and cPS \RvMJ{(with the spatial resolution, $ R=H/\mathit{dp}=200 $)} accurately estimate the impact load at the pressure measurement point, $ S $, and the wave propagation on the horizontal bed, \RvMJ{$ x_f $}, compared with the experimental results. The discrepancies between the numerical and experimental measurements are related to the air-cushioning effects neglected in the single fluid phase simulations and the solid boundary conditions \cite{Mokos2017}. \RvMJ{We should highlight that the local pressure extracted every 0.02 seconds may not reflect the high-frequency pressure fluctuations due to the weak compressibility of the fluid phase.}  

\begin{figure}[H]
	\centering
	\includegraphics[width=\textwidth]{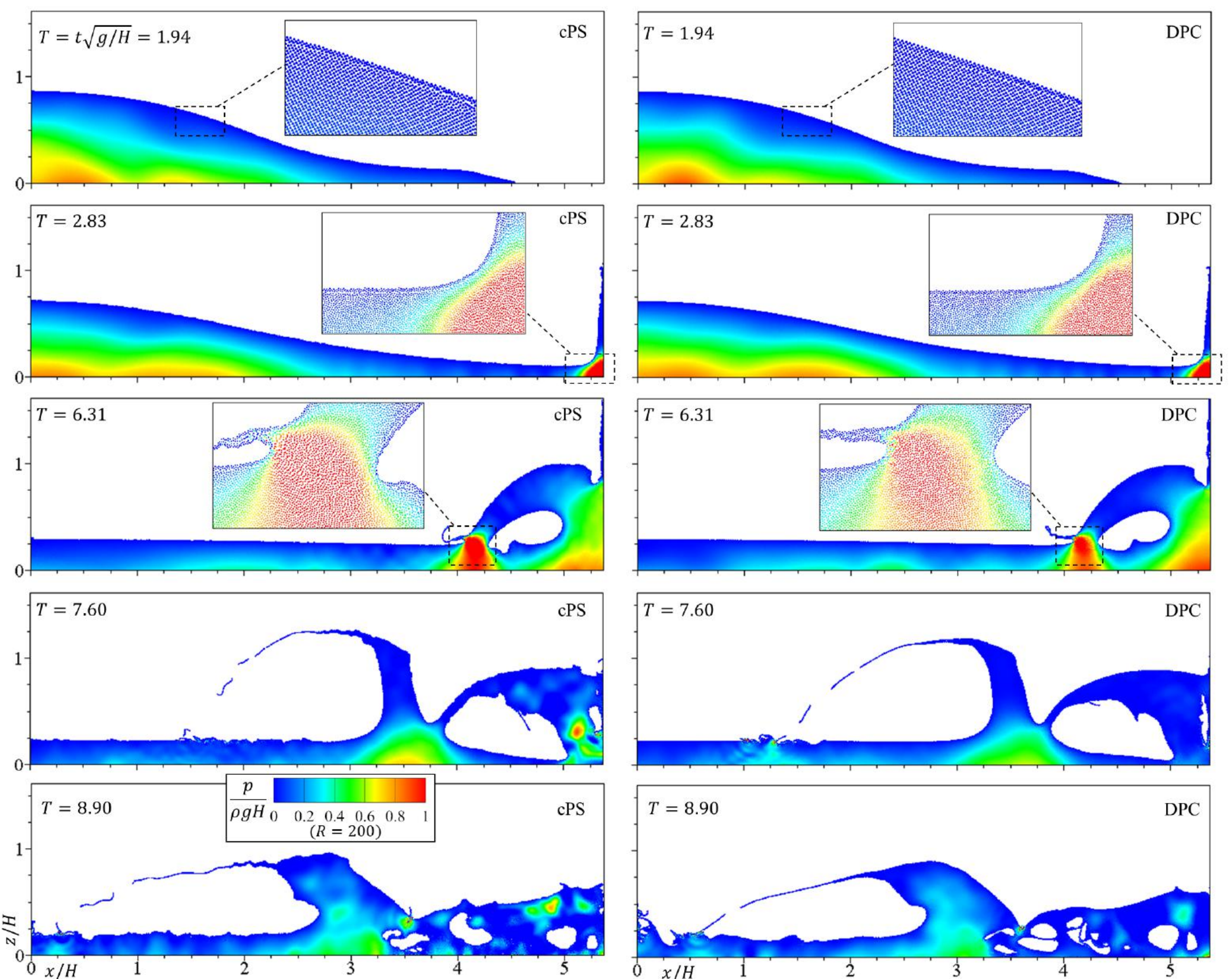}
	\caption{Dam break: flow evolutions and non-dimensional pressure fields with cPS and DPC (the left and right columns, respectively) with $ R=200 $}
	\label{fig_DBPr}
\end{figure}

\begin{figure}[H]
	\centering
	\includegraphics[width=\textwidth]{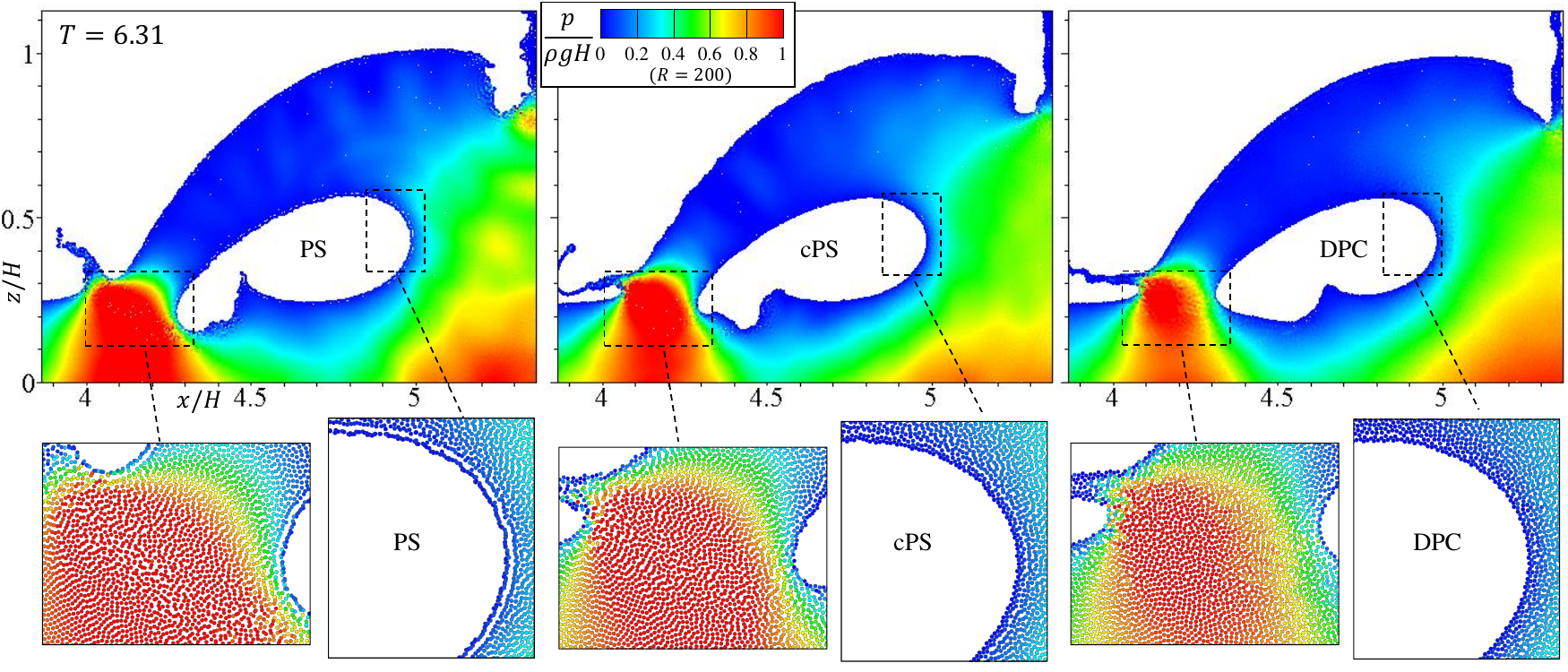}
	\caption{Dam break: the particle distribution and the non-dimensional pressure field as the wave impacts the free-surface at $ T=6.31 $ (simulated by PS, cPS and DPC techniques where $ R=200 $)}
	\label{fig_DBPrPS}
\end{figure}

Fig. \ref{fig_DBPr} represents the flow evolution and the non-dimensional pressure field simulated by cPS and DPC. The overall flow evolution with both approaches are almost identical, however, we observe that DPC reduces pressure fluctuations during highly dynamic flows and impact events at $ T=7.60 $ and $ 8.90 $. To illustrate the efficiency of the regularization techniques, we display snapshots of the plunging wave at $ T=6.31 $ (Fig. \ref{fig_DBPrPS}). The results shown confirm the effectiveness of DPC embedded into the PS method through equations (\ref{eq:cPS}) and (\ref{eq:cPS2}) (i.e., the cPS method) for improving the particle distribution at the impact and free-surface regions compared with PS. Moreover, DPC (i.e, equation (\ref{eq:DPC})) applied among all the fluid particles results in more uniform particle distribution without involving any boundary treatments (Fig. \ref{fig_DBPrPS}). 

\begin{figure}[H]
	\centering
	\includegraphics[width=\textwidth]{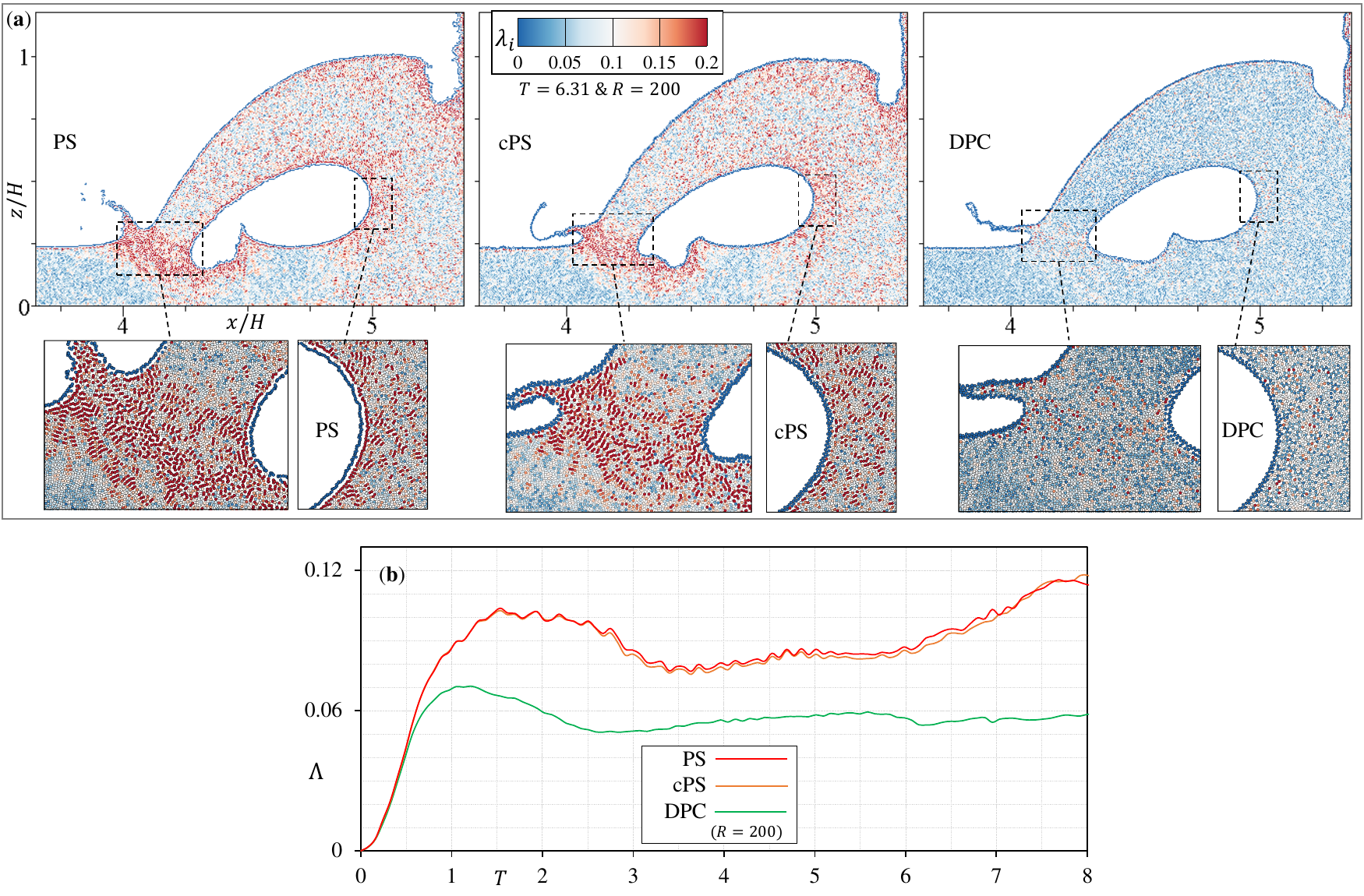}
	\caption{Dam break: (a) the local spatial particle disorder, $ \lambda_i $, \RvMJ{at $ T=6.31 $} and (b) the evolution of its global value, $ \Lambda $, with \RvMJ{PS,} cPS, and DPC techniques where $ R=200 $}
	\label{fig_DBLam}
\end{figure}

To quantify the improvements, we compare the local spatial particle disorder and the time evolution of its global value, i.e., $ \lambda_i $ and $ \Lambda $, respectively, simulated by \RvMJ{PS}, cPS, and DPC (Fig. \ref{fig_DBLam}). \RvMJ{cPS reduces spatial particle disorder at the vicinity of the free-surface compared to PS; yet, both PS and cPS manifest irregular particle distribution at the impact location as $ \lambda>0.2 $.} DPC proves to be capable of reducing the spatial particle disorder where the wave impacts the free-surface and pressure suddenly increases (Fig. \ref{fig_DBLam}-a). Also, the global particle disorder with DPC is considerably less than the models with \RvMJ{PS} and cPS, especially during the violent flow of the dam beak (Fig. \ref{fig_DBLam}-b); \RvMJ{with DPC, $ \Lambda $ remains less than 0.06, while with cPS and PS, $ \Lambda $ exceeds 0.11 at $ T=8 $}.

\begin{figure}[H]
	\centering
	\includegraphics[width=\textwidth]{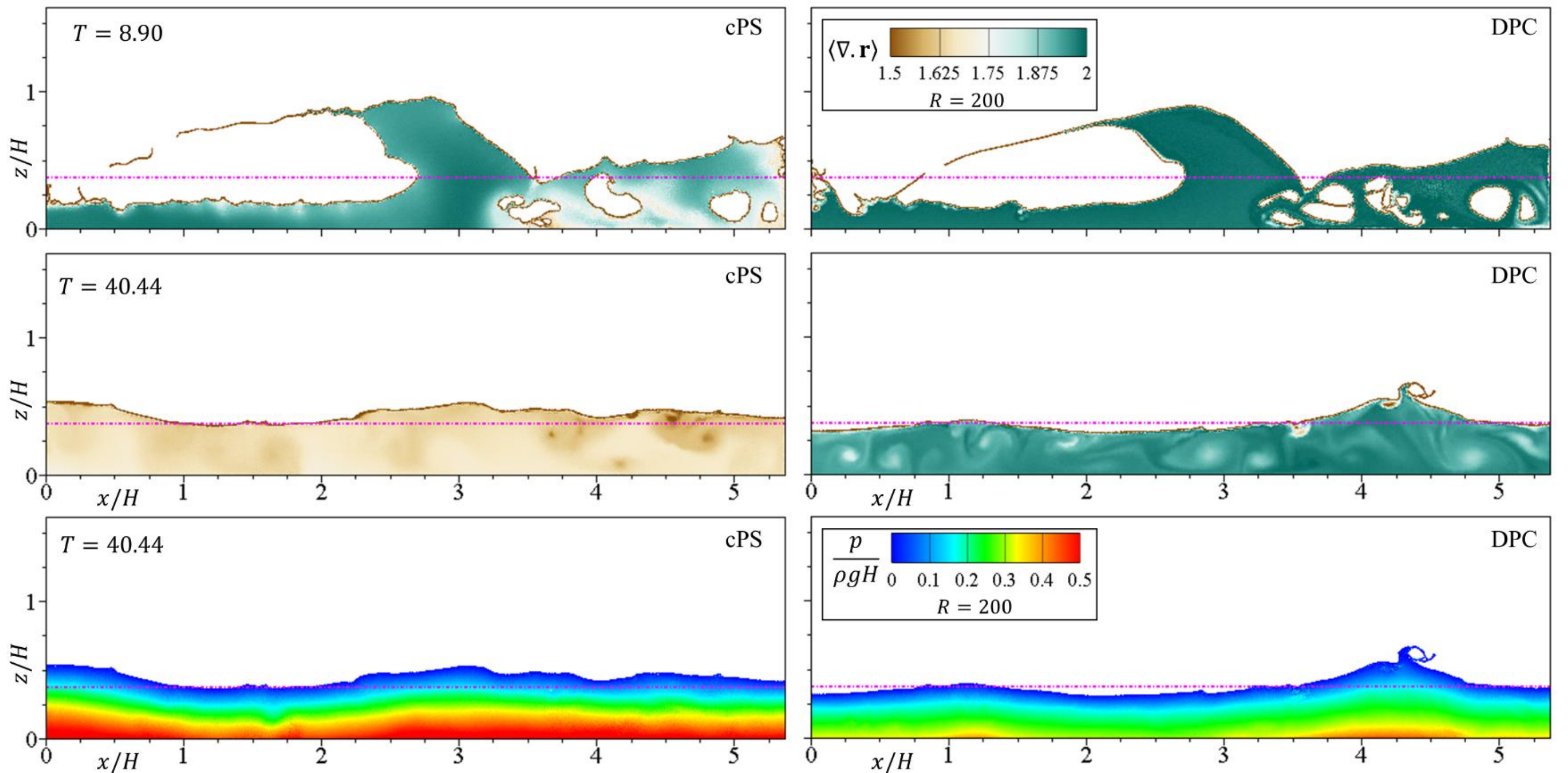}
	\caption{Dam break: $ \langle{\nabla\cdotp\mathbf{r}}\rangle_i $ and the non-dimensional pressure field at $ T=8.90 $ and the final stage, $ T=40.44 $, with cPS and DPC where $ R=200 $. The dash-dot line is the expected fluid height at the final equilibrium state.}
	\label{fig_DBdivr}
\end{figure}

\begin{figure}[H]
	\centering
	\includegraphics[width=\textwidth]{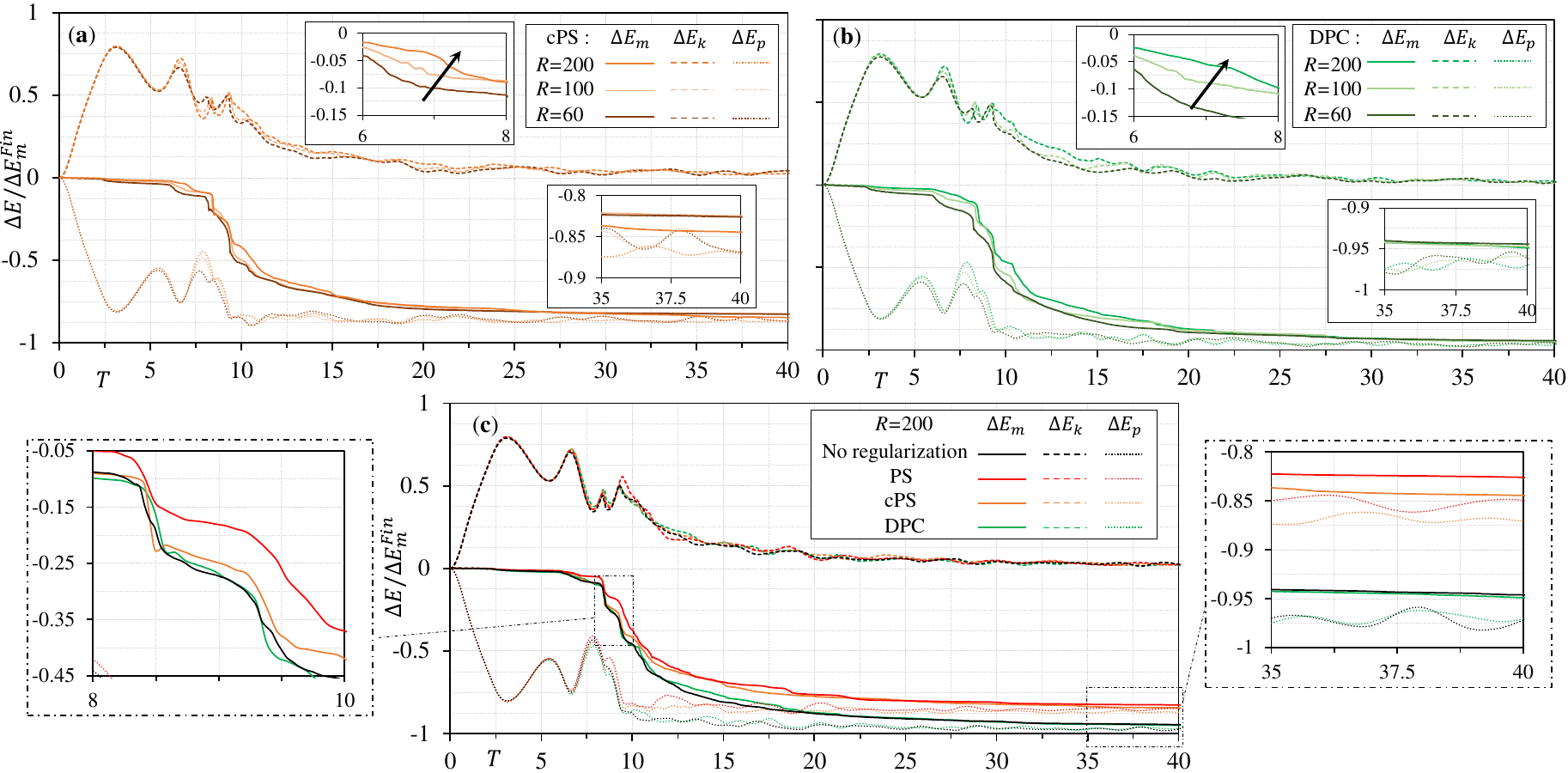}
	\caption{Dam break: evolution of the global energies by cPS and DPC with different spatial resolutions, $ R $, plotted in graphs (a) and (b), respectively. Graph (c) compares the energy evolution of the model with no particle regularization with the profiles of \RvMJ{PS,} cPS, and DPC where $ R=200 $.}
	\label{fig_DBEn}
\end{figure}

Next, we study the conservation of the fluid volume and the global energy evolution of the simulations. In a volume conservative model, the local value of $ \langle{\nabla\cdotp\mathbf{r}}\rangle_i $ is expected to remain equal to the space dimension of the test case (i.e., for this 2D case, we expect $ \langle{\nabla\cdotp\mathbf{r}}\rangle_i \cong 2 $). Fig. \ref{fig_DBdivr} manifests the unrealistic volume expansion of the fluid phase due to the shifting of the internal particles (as $ \langle{\nabla\cdotp\mathbf{r}}\rangle_i<1.75 $ in some regions at $ T=8.90 $ and over the entire fluid domain, at $ T=40.44 $); DPC does not suffer from this numerical issue (noting that $
\langle{\nabla\cdotp\mathbf{r}}\rangle_i > 1.8$ over the main fluid domain). We also observe that the unphysical volume expansion diverges the hydrostatic pressure expected in the late stages of the flow (at $ T=40.44 $ shown in Fig. \ref{fig_DBdivr}). Furthermore, we plot the energy evolution of the system normalized by $ \Delta E^{fin}_m=E^0_p-E^{\infty}_p $ in Fig. \ref{fig_DBEn} (where $ E^{0}_p $ is the initial potential energy and $ E^{\infty}_p $ is the final expected potential energy with the fluid flow reaching the equilibrium state in the rectangular tank). For both cPS and DPC, increasing $ R $ reduces the dissipation of the mechanical energy during the main impact events showing the numerical convergence of the results (Fig. \ref{fig_DBEn}-a \& b). DPC estimates accurate evolution of the global energy (i.e., $  \Delta E_p/ \Delta{E^{fin}_m} \cong -0.975 $ as ideally should reach -1.0). The energy evolution with DPC is almost identical to the results of the model with no particle regularization (Fig. \ref{fig_DBEn}-c); thus, DPC effectively improves the particle distribution without manipulating the global flow properties. In contrast, the unphysical volume expansion with the particle shifting \RvMJ{(in either PS or cPS forms)} increases the potential energy of the system not reaching the expected value (i.e., $  \Delta E_p/ \Delta{E^{fin}_m} \cong -0.875 $ \RvMJ{and $ -0.85 $ with cPS and PS, respectively}). \RvMJ{Particularly during the violent flow deformations (i.e., $ T\cong8-10 $), the standard PS excessively increases the global mechanical energy of the system (shown in the zoomed view of graph (c) in Fig. \ref{fig_DBEn}).}

\begin{figure}[H]
	\centering
	\includegraphics[width=\textwidth]{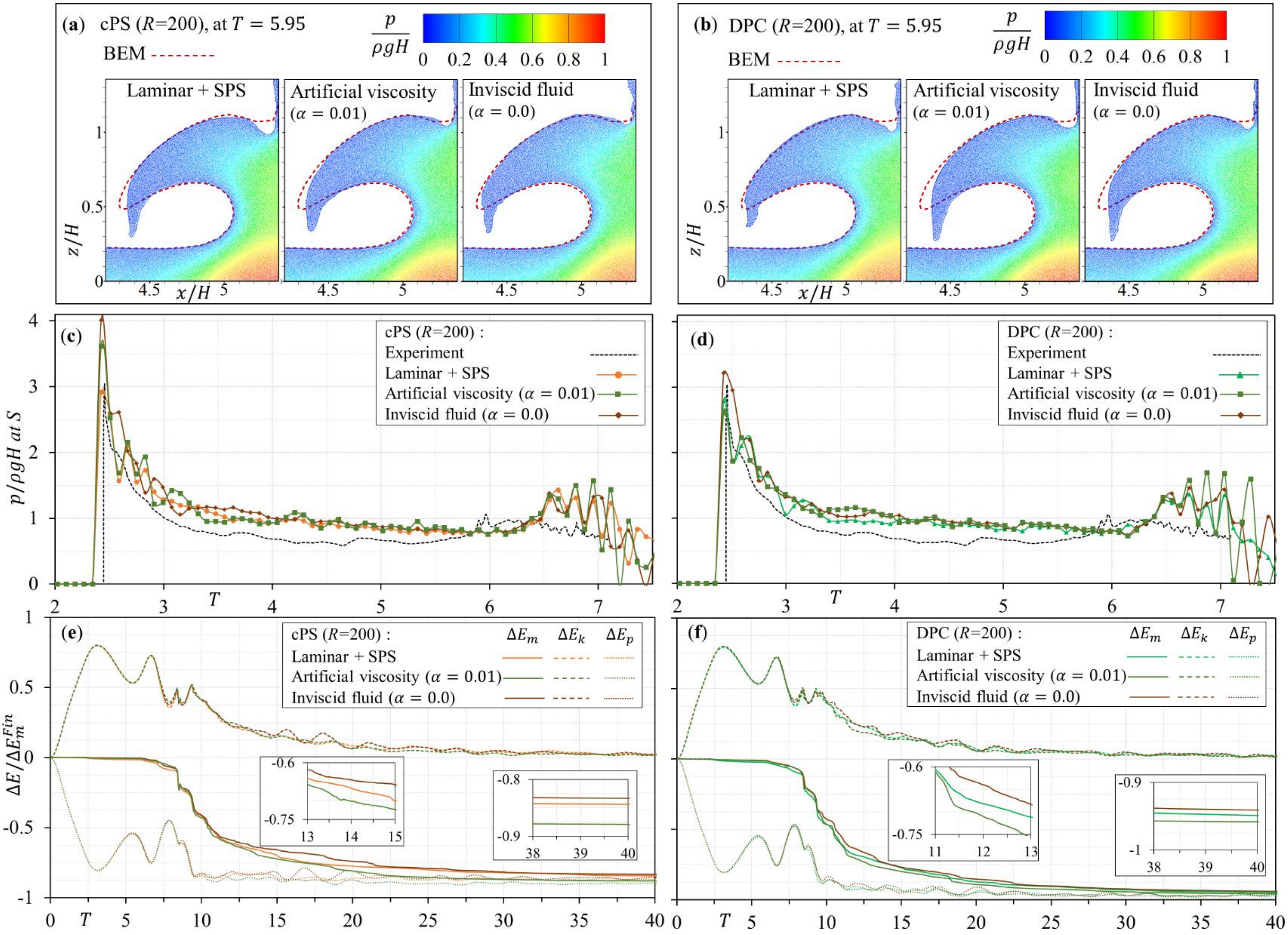}
	\caption{\RvMJ{Dam break: flow evolution and the non-dimensional pressure fields (a, b), the impact load, $ p $, at \textit{S} (c, d), and the energy evolution, $ \Delta E $, (e, f) with different viscosity models, i.e., the laminar+SPS model, the artificial viscosity term ($ \alpha=0.01 $), and the inviscid fluid ($ \alpha=0$), simulated by cPS and DPC where $ R=200 $. The dashed red line in (a) and (b) is the free surface at $ T=5.95 $ predicted by the BEM solver from \cite{Greco2004}. The dashed black line in (c) and (d) is the experimental measurement of the impact load from \cite{Lobovsky2014}.}}
	\label{fig_DBVisco}
\end{figure}

\RvMJ{To investigate the role of the shear force term in the pressure and flow evolution, we compare the results of the dam-break case simulated by cPS and DPC (with $ R=200 $) considering different viscosity models, i.e., the Laminar+SPS model (equation (\ref{eq:LamSPS})), the artificial viscosity term (see \cite{Dominguez2021}), and with no viscosity (as if water is an inviscid fluid). A non-dimensional coefficient, $ \alpha $, adjusts the intensity of the artificial viscosity \cite{Monaghan&Gingold1983, Dominguez2021}. We implement the artificial viscosity term by setting $ \alpha$ to 0.01 and considering the slip boundary condition at the solid walls. To simulate water as an inviscid fluid, $ \alpha $ is set to 0.0. As illustrated in Fig. \ref{fig_DBVisco}- (a) and (b), in all the cases, the pressure field is smooth, and the free surface match quite well with the reference solution (the dashed red line reported by Greco et al. \cite{Greco2004} using a Boundary Element Method (BEM)) at $ T=5.95 $. Fig. \ref{fig_DBVisco}- (c) and (d) show that the averaged local pressure of all the viscosity models evolves similarly during the impact events and is compatible with the experimental measurement (the dashed black line from \cite{Lobovsky2014}). These results confirm that the smoothness of the pressure field and the evolution of the estimated impact load (with the adopted density diffusion term (\ref{eq:Di}) and the cPS or DPC technique) are almost independent of the viscosity formulation. Moreover, Fig. \ref{fig_DBVisco}- (e) and (f) indicate that the Laminar+SPS formulation (in the form implemented in DualSPHysics) dissipates slightly less mechanical energy in comparison with the artificial viscosity term. Accordingly, we choose the Laminar+SPS model for the shear force calculation to evaluate the effect of particle regularization techniques on the system's conservation properties without incorporating energy dissipation of the artificial term.}

\subsection{2D water sloshing in a tank}
\begin{figure}[H]
	\centering
	\includegraphics[width=\textwidth]{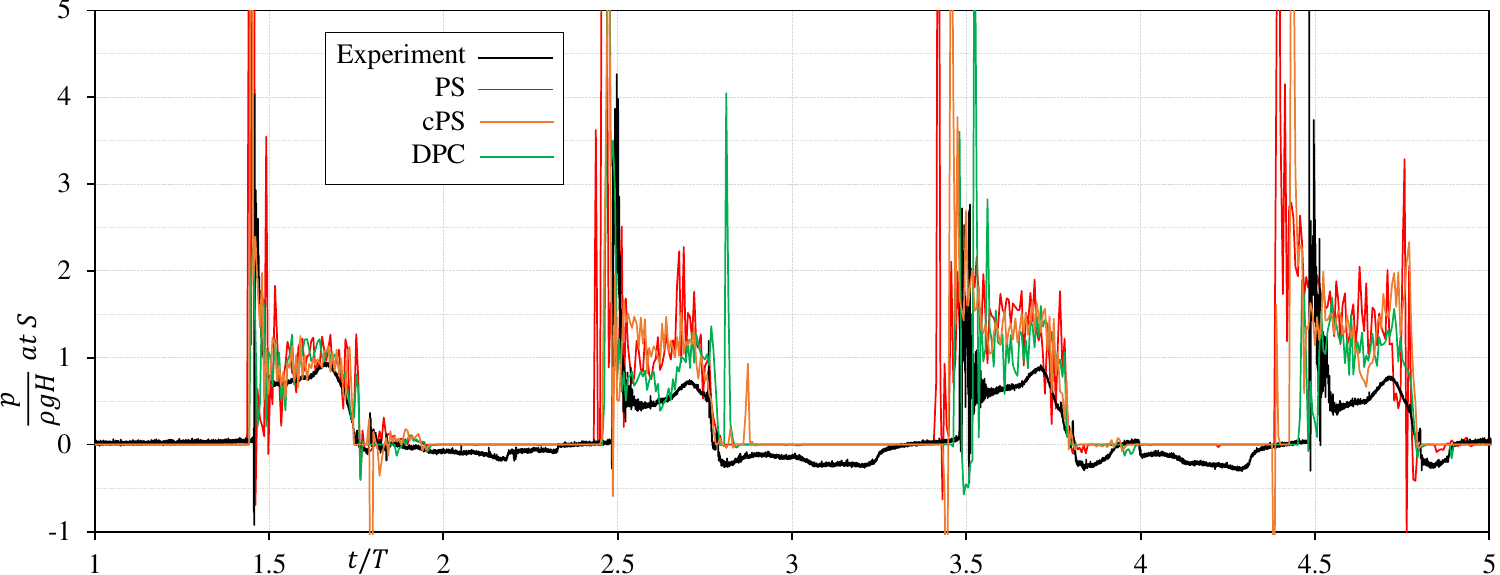}
	\caption{Water sloshing: the local averaged pressure obtained by \RvMJ{PS,} cPS, and DPC ($ dp=0.002$ ($ m $)) compared with the experimental measurements at the probe location, $ S $, reported by Souto-Iglesias et al. \cite{Souto-Iglesias2015}. \RvMJ{Numerical results are extracted every 0.01 seconds.} }
	\label{fig_SLexp}
\end{figure}

\begin{figure}[H]
	\centering
	\includegraphics[width=\textwidth]{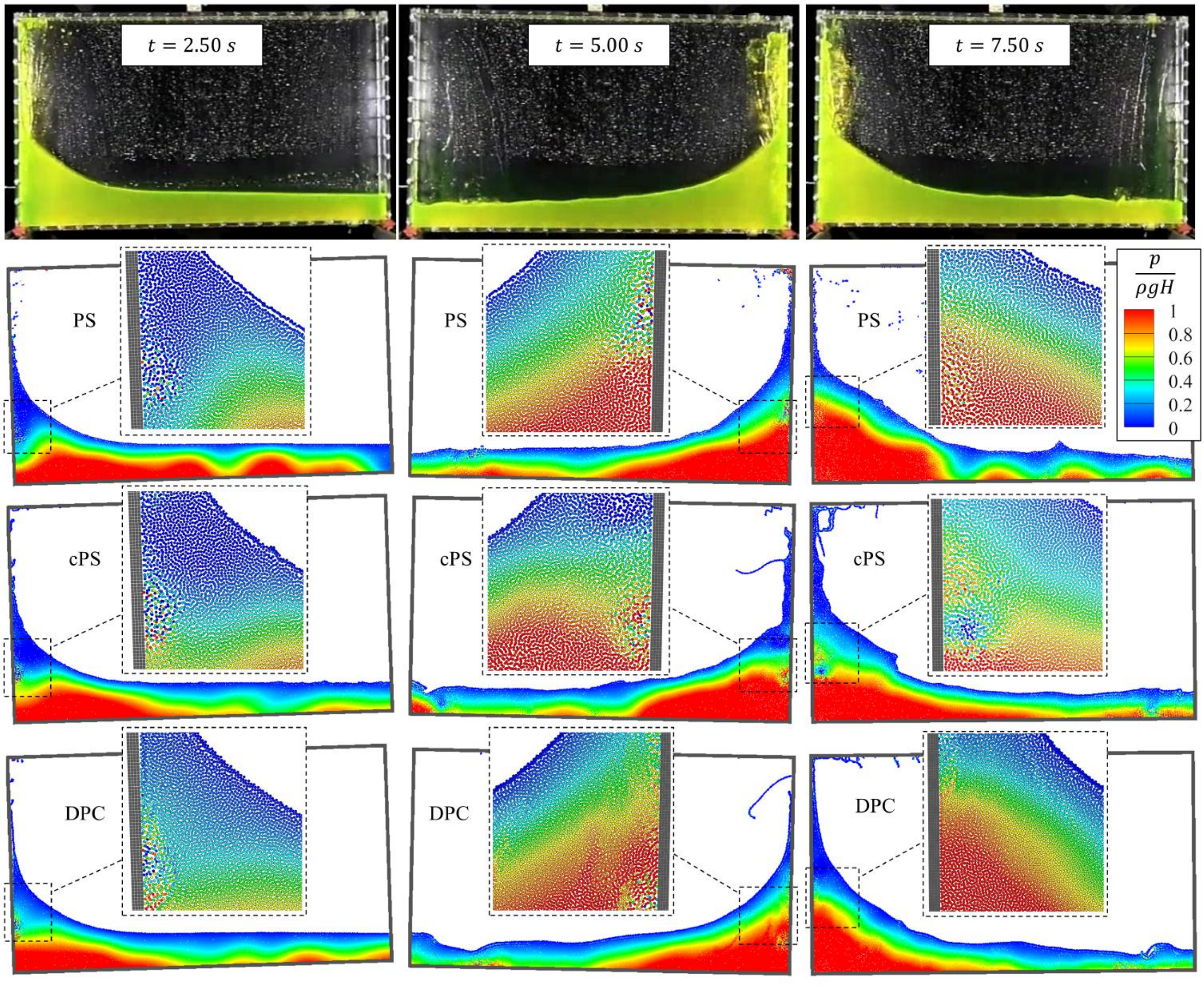}
	\caption{Water sloshing: the flow evolutions and the non-dimensional pressure fields with \RvMJ{PS,} cPS, and DPC ($ dp=0.002$ ($ m $)) compared with snapshots of the experiment on the top-row (taken from \cite{Souto-Iglesias2015})}
	\label{fig_SLPress}
\end{figure}

Here, we \RvMJ{present} and discuss the numerical simulations of the 2D water sloshing and its lateral water impacts. The time history of the impact loads at the location of the pressure sensor, \textit{S}, is compared with the experimental measurements by \cite{Souto-Iglesias2015}. \RvMJ{Fig. \ref{fig_SLexp} shows that while PS and cPS slightly overestimate the local averaged pressure (at $ t/T\cong2.6, 3.7, 4.75 $), DPC predicts more accurate impact loads with less pressure noises}. The remaining incompatibility between the numerical and experimental results is due to the fact that the single phase simulation ignores the air-cushioning effects. \RvMJ{Also, the periodic rotation of the solid walls may affect the accuracy of the estimated impact load increasing local pressure fluctuations observed in Fig. \ref{fig_SLexp} \cite{English2021}}. Fig. \ref{fig_SLPress} illustrates snapshots of the flow evolution (taken from the experiment represented by \cite{Souto-Iglesias2015}) and the non-dimensional pressure fields with \RvMJ{PS,} cPS, and DPC. Although all the models capture the overall flow evolution, yet, DPC results in isotropic and more compact particle distributions during the lateral impacts (see the zoom-in sections of Fig. \ref{fig_SLPress}). \RvMJ{With the extreme kernel truncation at the free surface, the standard approximation operators diverge the pressure and velocity fields affecting the dynamic free-surface boundary condition \cite{You2021}. The numerical errors due to the incomplete kernel support with no special treatment of PS near the free surface lead to dominant particle clustering. DPC and cPS improve the particle distribution in this region; nevertheless, to further resolve the accumulation of particles at the free-surface vicinity, one can implement high-order discretization operators (e.g., the high-order density diffusion terms \cite{Antuono2010, Jandaghian2021_AWR}) and virtual background nodes (e.g., the virtual particle technique by Duan et al. \cite{Duan2017} and the Background Mesh scheme by You et al. \cite{You2021}).}

\begin{figure}[H]
	\centering
	\includegraphics[width=\textwidth]{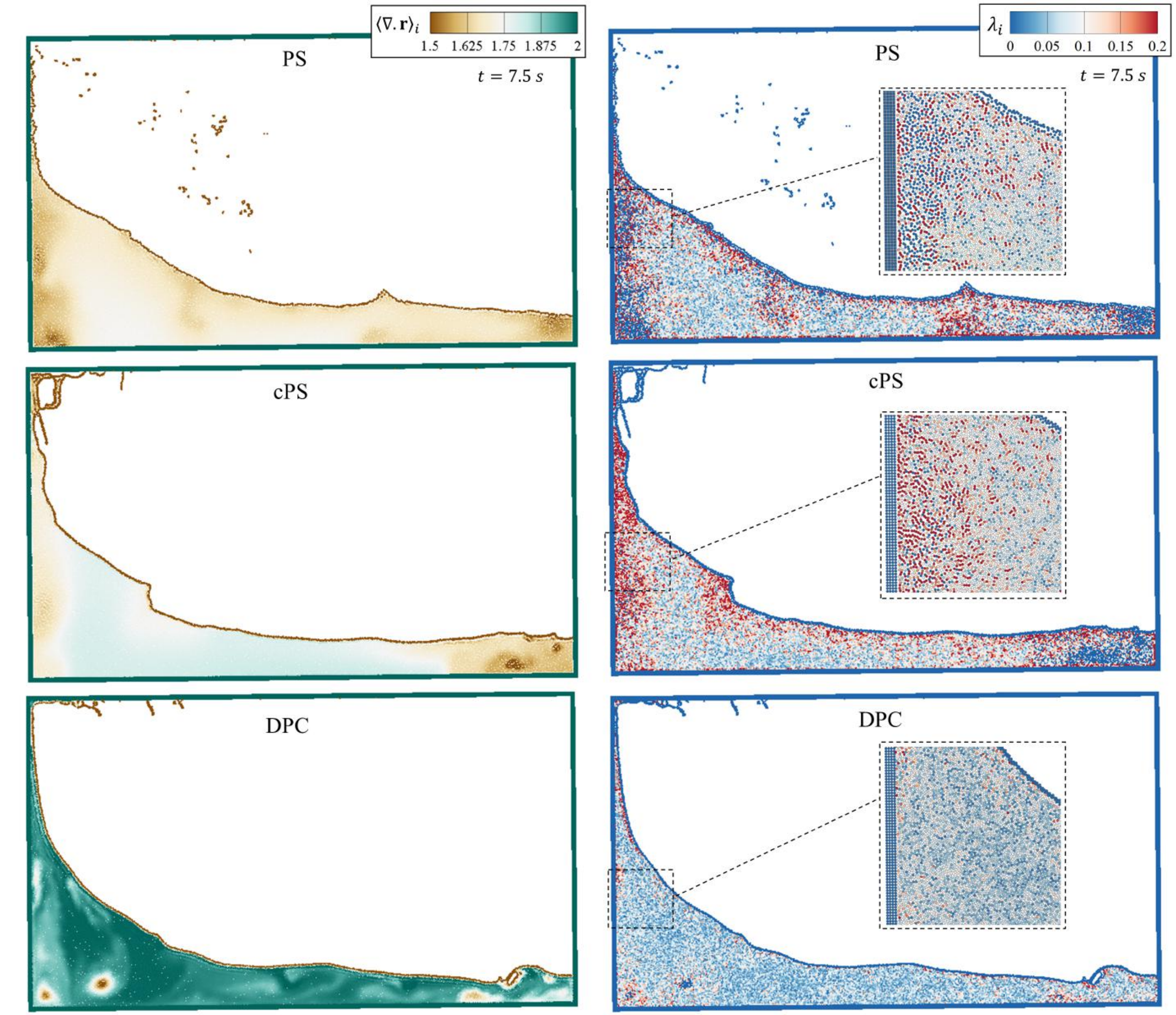}
	\caption{Water sloshing: $ \langle{\nabla\cdotp\mathbf{r}}\rangle_i $ and the local spatial particle disorder, $ \lambda_i $ at $ t=7.50 $ seconds where $ dp=0.002$ ($ m $)}
	\label{fig_SLdivL}
\end{figure}

To quantify the effectiveness of the models, we represent the estimated value of $ \langle{\nabla\cdotp\mathbf{r}}\rangle_i $ and the local spatial disorder of particles, $ \lambda_i $ (in Fig. \ref{fig_SLdivL}). DPC retains $ \langle{\nabla\cdotp\mathbf{r}}\rangle_i>1.8$ (which is expected te be $ \cong 2$); due to the unphysical volume expansion by \RvMJ{PS and} cPS, $ \langle{\nabla\cdotp\mathbf{r}}\rangle_i<1.75 $ becomes dominant over the entire fluid domain. Moreover, DPC improves uniform particle distribution in comparison to \RvMJ{PS and} cPS by which $ \lambda_i $ increases to more than 0.2 where the lateral impact occurs at $ t=7.50 $ seconds. \RvMJ{We also observe that, with PS and cPS, the volume expansion and the extreme particle clustering issue within the interior domain of the flow make conditions (\ref{eq:bi}) incapable of accurately detecting the boundary and splashed particles (noting that in some internal regions $ \lambda_i $ is incorrectly set to zero).} 

\begin{figure}[H]
	\centering
	\includegraphics[width=\textwidth]{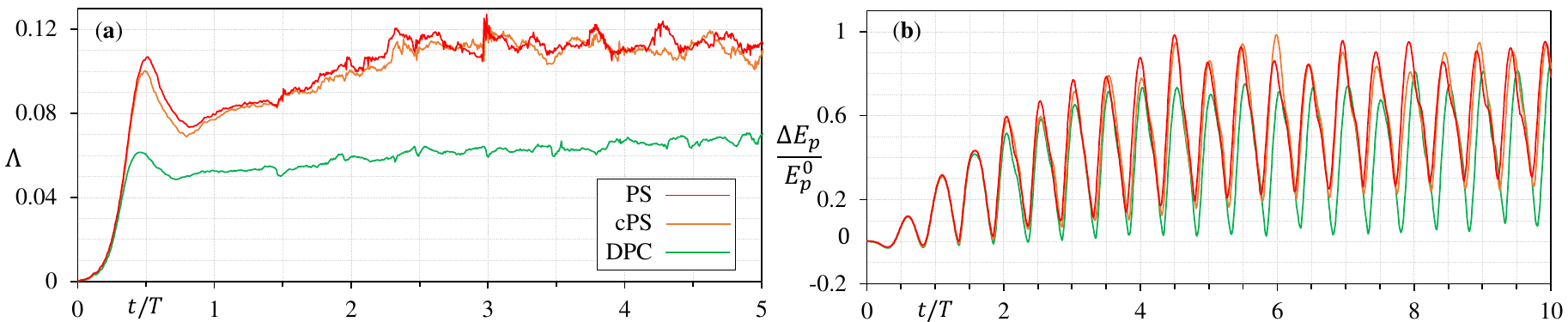}
	\caption{Water sloshing: the time evolution of the global spatial particle disorder, $ \Lambda $, and the global potential energy, $ E_p $, where $ dp=0.002$ ($ m $) (represented in graphs (a) and (b), respectively)}
	\label{fig_SLGLEn}
\end{figure}

Next, we plot the time evolution of the global particle disorder, $ \Lambda $, and the global potential energy, $ E_p $, in Fig \ref{fig_SLGLEn}-a and b, respectively. The potential energy is normalized by its initial global value, $ E^0_p $. After 5 cycles of rotation, $ \Lambda $ increase to more that 0.11 by \RvMJ{PS and} cPS, while DPC keeps $ \Lambda $ less than 0.07 showing more regular particle distribution over the entire domain. We again observe that, unlike DPC, \RvMJ{PS and} cPS diverge the minimum potential energy of the system in the long-term simulation of this violent free-surface flow. This divergence of potential energy, manifested as the unphysical volume expansion, originates from the inconsistent implementation of the shifting equation (i.e., without considering the additional diffusion/cohesion terms of the particle shifting transport-velocity \cite{Jandaghian2021_JCP, Sun2019, Hong-Guan2021}).

\label{sec:2DST}

\subsection{3D water dam-break against a rigid obstacle}
\label{sec:3DBO}
\begin{figure}[H]
	\centering
	\includegraphics[width=\textwidth]{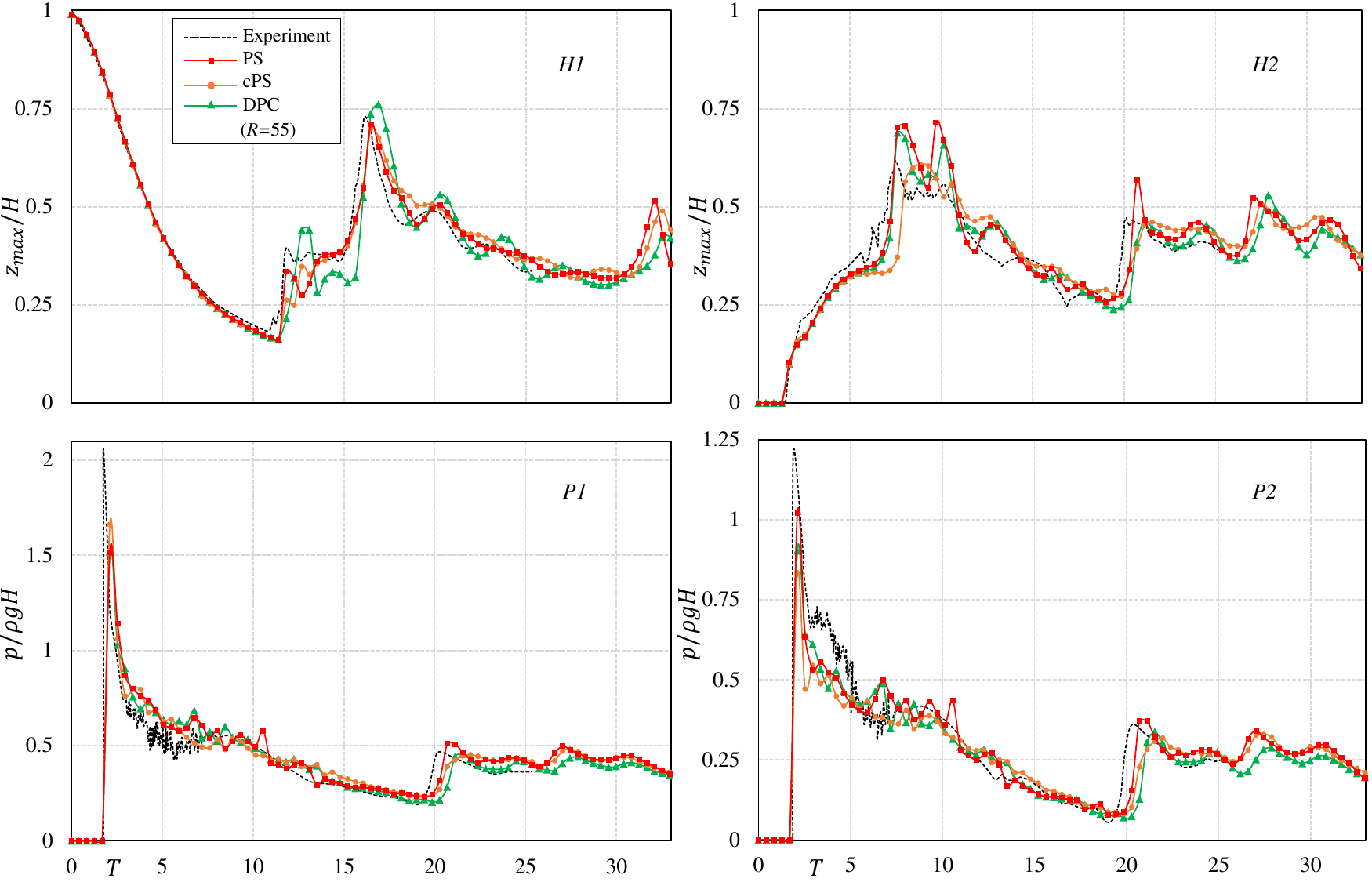}
	\caption{3D dam break: the fluid height at \textit{H1} and \textit{H2}, and the local pressure at \textit{P1} and \textit{P2} by \RvMJ{PS,} cPS, and DPC (where \RvMJ{$ R=H/\mathit{dp}=55 $}) validated with the experimental measurements by Kleefsman et al. \cite{Kleefsman2005}. \RvMJ{Numerical results are extracted every 0.1 seconds.}}
	\label{fig_3DBExp}
\end{figure}

In Fig. \ref{fig_3DBExp}, by plotting the extracted numerical results against the experimental measurements (by \cite{Kleefsman2005}), we show that the implemented models predict the fluid flow evolutions and the impact loads on the rigid obstacle. The temporal evolution of the fluid heights over \textit{H1} and \textit{H2} lines and the local pressures on \textit{P1} and \textit{P2} points (with \RvMJ{PS,} cPS, and DPC, \RvMJ{where the spatial resolution, $ R=H/\mathit{dp}=55 $}) are compatible with the experimental results. Similar to the other single fluid phase simulations (e.g., \cite{Jandaghian2021_JCP, English2021, Sato2021}), a delay occurs in the impact load of the returning wave (at $ T\cong20 $). 

\begin{figure}[H]
	\centering
	\includegraphics[width=\textwidth]{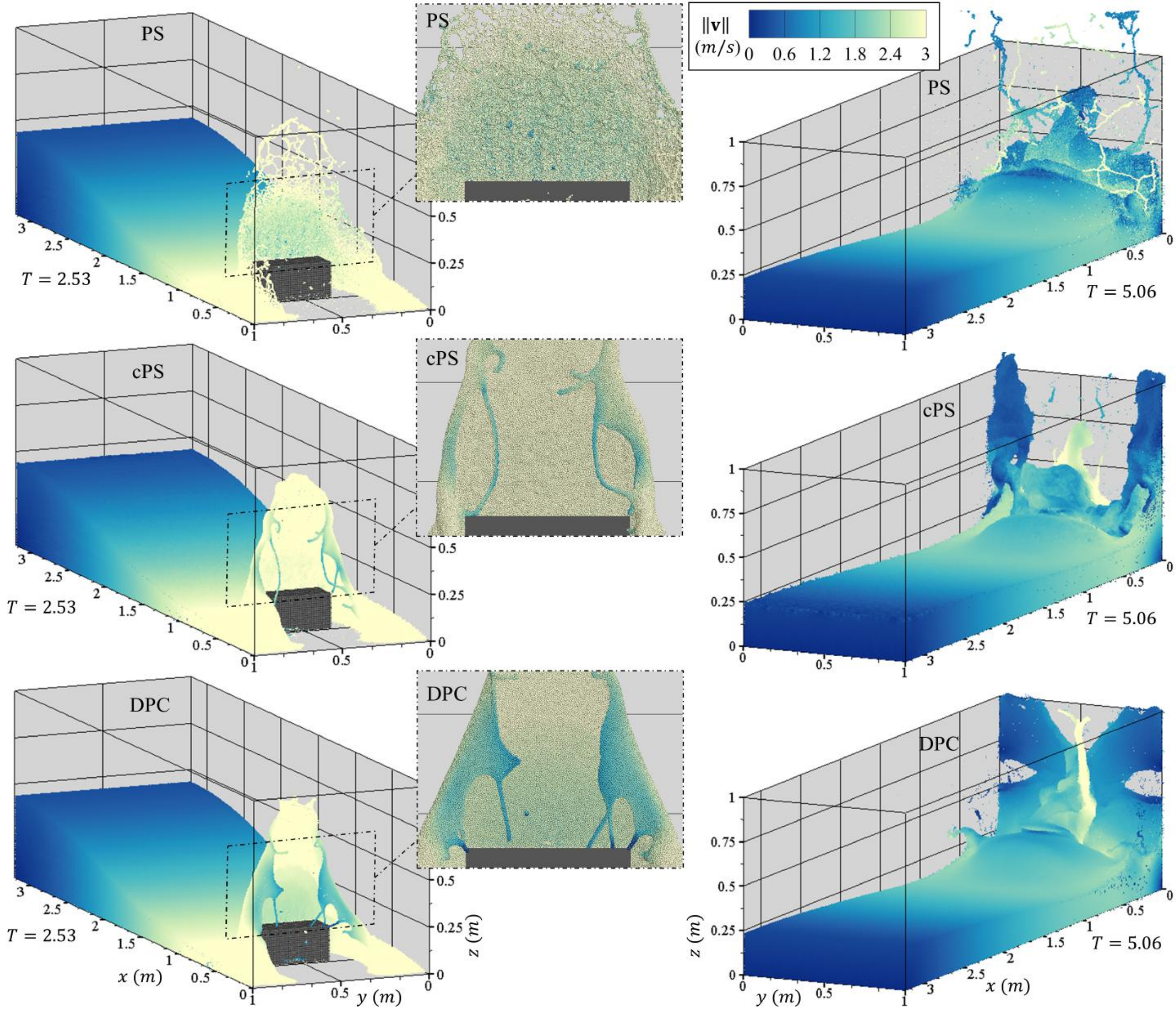}
	\caption{3D dam break: the flow evolution and the magnitude of velocity by \RvMJ{PS,} cPS, and DPC (the top, middle and bottom rows, respectively) at $ T=2.53 $ and $ 5.06 $, with $ R=110 $.}
	\label{fig_3DBVel}
\end{figure}

\begin{figure}[H]
	\centering
	\includegraphics[width=\textwidth]{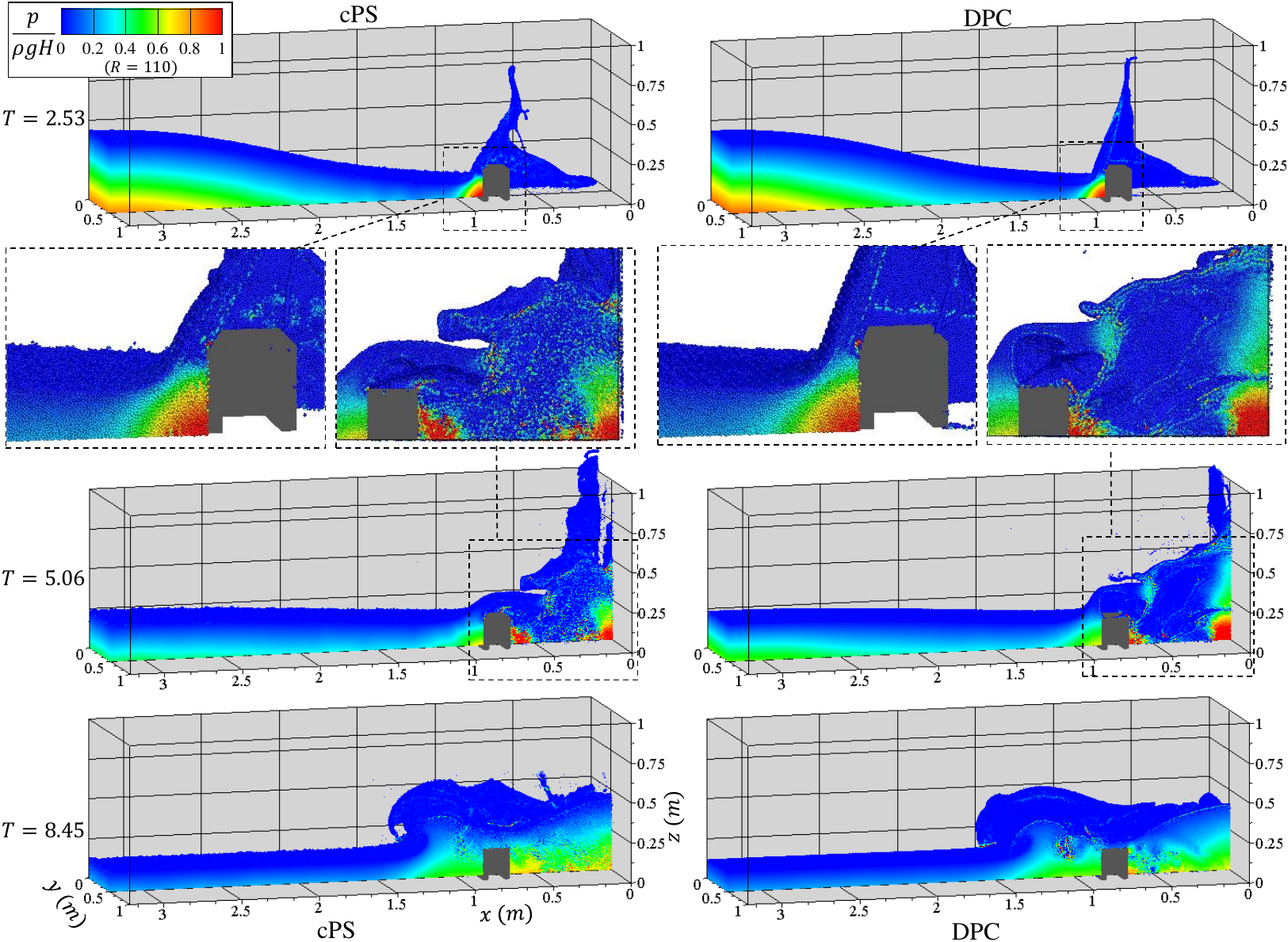}
	\caption{3D dam break: non-dimensional pressure field at the middle-section of the reservoir, $ y=0.5 $ ($ m $), by cPS and DPC (the left and right columns, respectively) at $ T=2.53 $, $ 5.06 $, and $ 8.45 $, with $ R=110 $.}
	\label{fig_3DBPress}
\end{figure}

\begin{figure}[H]
	\centering
	\includegraphics[width=\textwidth]{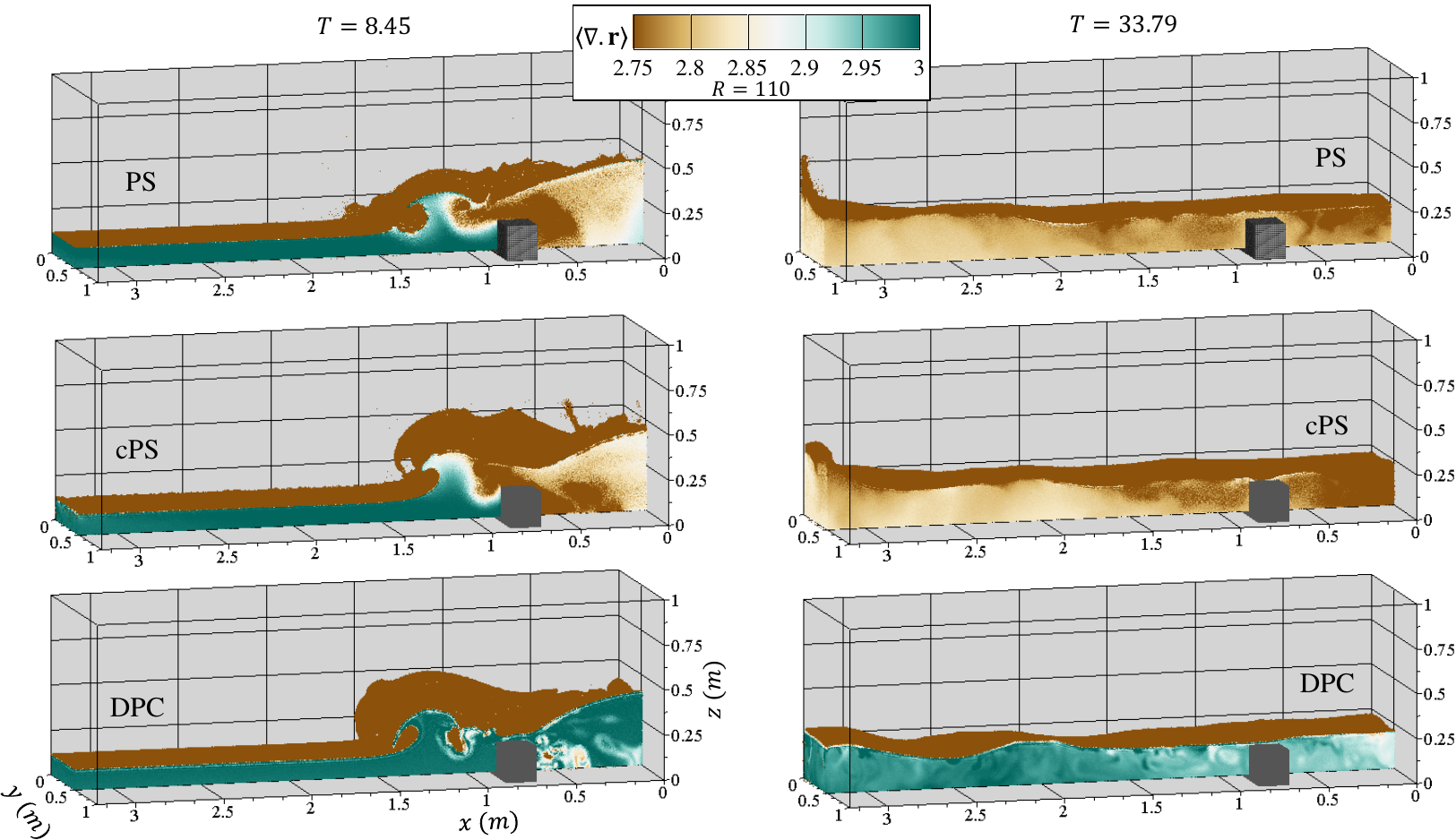}
	\caption{3D dam break: $ \langle{\nabla\cdotp\mathbf{r}}\rangle_i $ at the middle-section of the reservoir, $ y=0.5 $ ($ m $), by \RvMJ{PS,} cPS, and DPC (the top, middle, and bottom rows, respectively) at $ T=8.45 $ and $ 33.79 $, with $ R=110 $.}
	\label{fig_3DBDivR}
\end{figure}

Fig. \ref{fig_3DBVel} illustrates the violent free-surface deformations and the magnitude of the velocity field as the water flow impacts the obstacle (at $ T\cong2.5 $), and further, climbs up the front wall of the reservoir (at $ T=5.06 $). \RvMJ{The standard PS cannot eliminate fluid fragmentation due to highly dynamic flow deformations (shown in the zoomed view of the splashed flow); however, cPS and DPC simulate more uniform flow evolution.} Fig. \ref{fig_3DBPress} displays the pressure field at the middle-section of the tank (where $ y=0.5 $ ($ m $)) simulated by cPS and DPC. During the formation of the submerged waves, DPC shows slightly less pressure noises (e.g., at $ T=5.06 $ and $ 8.45 $). Moreover, the inconsistent PS models \RvMJ{(in the PS and cPS forms)} clearly result in the volume-non-conservation issue as $ \langle{\nabla\cdotp\mathbf{r}}\rangle_i<2.87 $ in the internal regions of the fluid flow (shown in Fig. \ref{fig_3DBDivR}, at $ T=7.45 $ and $ T=33.79 $). The divergence of $ \langle{\nabla\cdotp\mathbf{r}}\rangle_i$ highlights that employing a variable shifting coefficient ($ {D^F_i} $) and limiting the magnitude of the shifting vector are numerically insufficient for avoiding the unphysical volume expansion by the PS equation. This places emphasis on the implementation of the consistent PS algorithm (i.e., adopting the additional diffusive terms due to the shifting transport-velocity and the special free-surface treatments, e.g. in \cite{Sun2017_dplus, Jandaghian2021_JCP}) being essential for the long-term simulation of such complex free-surface flow. On the other hand, the results confirm that DPC, with being exempted from the complex boundary treatments and a conservative formulation, avoids numerical instabilities eliminating the particle clustering issue and representing smoother pressure fields.

\begin{figure}[H]
	\centering
	\includegraphics[width=\textwidth]{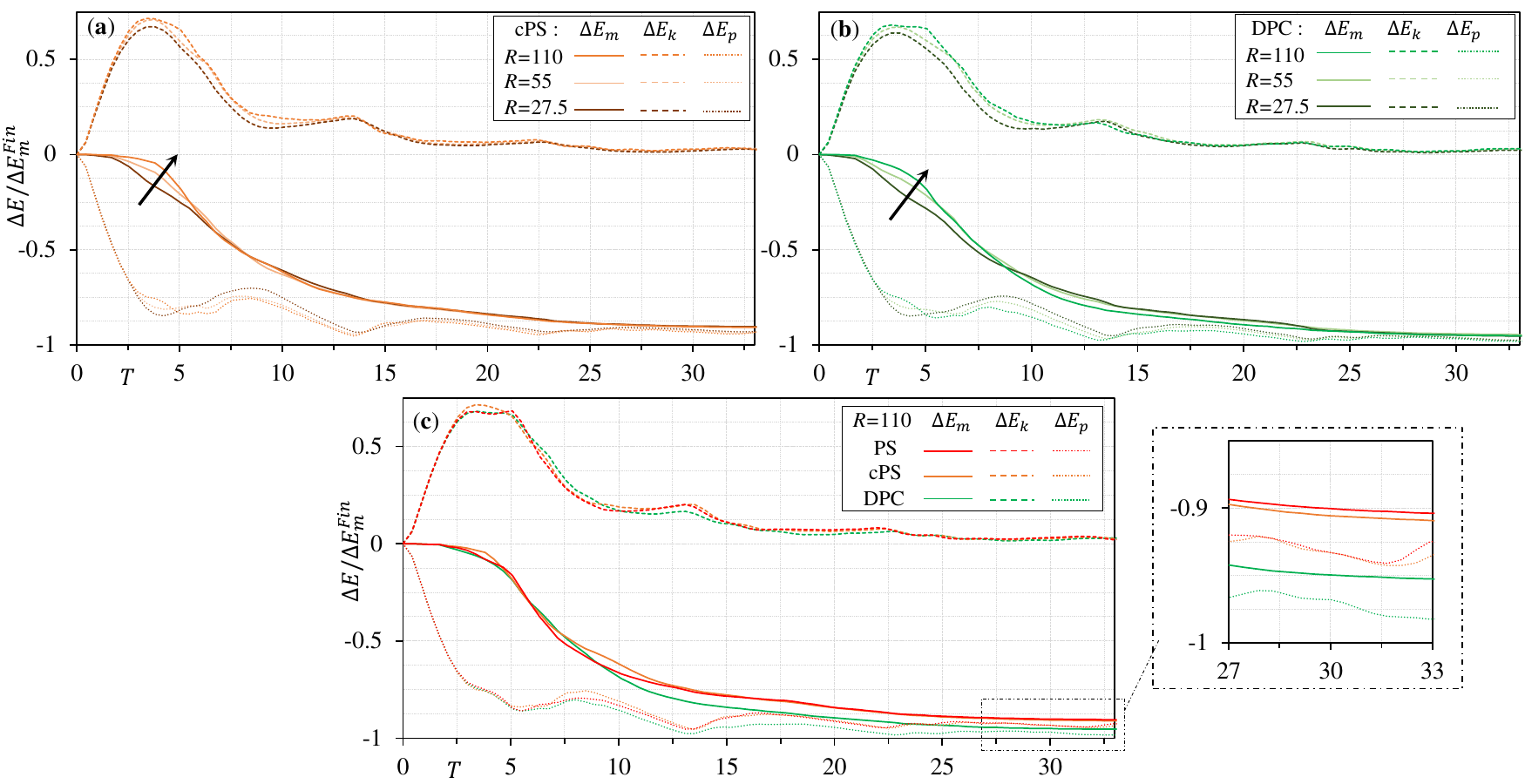}
	\caption{3D dam break: evolution of the energy components by cPS and DPC with different spatial resolutions ((a) and (b) graphs, respectively). Graph (c) compares the energy evolution by \RvMJ{PS,} cPS, and DPC where $ R=110 $.}
	\label{fig_3DBEn}
\end{figure}

Fig. \ref{fig_3DBEn} compares the time evolution of system's global energies. Increasing the spatial resolution reduces the energy dissipation during the impact events showing the numerical convergence of the simulations with either cPS or DPC (Fig. \ref{fig_3DBEn}-a \& b). Fig. \ref{fig_3DBEn}-c compares the energy evolutions where $ R=110 $. While DPC predicts the expected final potential energy (i.e., $  \Delta E_p/ \Delta{E^{fin}_m} \cong -0.975 $ as ideally should reach -1.0) the increase in the potential energy by PS (which starts from $ T\cong7.5 $ and remains over the simulation) evidences its inability for dealing with complex flow deformations. 

Finally, we evaluate the efficiency of the new implementations in DualSPHysics by comparing the simulation runtime with different regularization techniques (Tables \ref{table:DB-ti} and \ref{table:DB-ts}). The runtime per physical second and per iteration are denoted as $ t_\mathit{s} $ and $ t_\mathit{iter.} $, respectively. We simulate all the test cases using an NVIDIA Tesla V100 PCIe device (see the GPU specifications in Table \ref{table:GPU}). In this 3D problem, the number of fluid particles would be $ (79,380) $, $ (652,212) $, and $ (5,314,295) $ where  $ R=27.5,$ $ 55, $ and $ 110 $, respectively. In Tables \ref{table:DB-ti} and \ref{table:DB-ts}, $ (t_\mathit{(.)}/t^{PS}_\mathit{(.)}-1)\times100 $ gives the speed-up of runtime (in percent) with respect to the original model that implements the standard PS technique. Table \ref{table:DB-ti} shows that the implemented modifications (associated with the DPC and cPS formulations) increase the simulation runtime per iteration, $ t_\mathit{iter.} $, by 2-6.5 \%. Nevertheless, the numerical stability achieved by the proposed regularization techniques allows the model to adopt larger time steps as the total runtime, $ t_\mathit{s} $, reduces by 2.5-4.5 \% and 6-8.5 \% by cPS and DPC, respectively. \RvMJ{The time step of calculation is updated after every time step according to the CFL condition as a function of the maximum viscous diffusion and acceleration of particles \cite{Dominguez2021}. DPC eliminates unphysical high values of particle acceleration (at the impact events where pressure increases suddenly and inter-particle penetration occurs), and therefore, reduces the number of iterations over the simulation.} Overall, this reduction of simulation runtime (especially with DPC) justifies the implementation of effective particle regularization techniques not only for improving the numerical stability and accuracy, but also for increasing the efficiency of the computations.

\begin{table}[H]
	\caption{GPU specifications\strut} 
	\centering
	\small
	\begin{tabular}{c c} 
		\hline
		\multicolumn{2}{c}{NVIDIA Tesla V100 PCIe} \\
		\hline
		CUDA compatibility& 7.0 \\ 
		CUDA cores &  5120\\
		Multiprocessors &  80\\
		Global memory (MB) &  16160 \\
		GPU Maximum clock rate (MHz)&  1380 \\
		Memory clock rate (MHz)& 877\\
		Memory bus width (bits) & 4096\\
		\hline
	\end{tabular}
	\label{table:GPU}
\end{table}

\begin{table}[H]
	\caption{3D dam break: the simulation runtime per iteration denoted as $ t_\mathit{iter.} $. In this 3D problem, the number of fluid particles would be $ (79,380) $, $ (652,212) $, and $ (5,314,295) $ where  $ R=27.5,$ $ 55, $ and $ 110 $, respectively.\strut} 
	\centering
	\small
	\begin{tabular}{p{2.2cm} |c| c| c| c| c| c} 
		\hline
 	    \multirow{2}{1.8cm}{Model} & \multicolumn{3}{|c|}{$  t_\mathit{iter.}/10^{-3}$ (seconds)} & \multicolumn{3}{|c}{$ (t_\mathit{iter.}/t^{PS}_\mathit{iter.}-1)\times100 $}\\
 	    \cline{2-7} 
		& \multicolumn{1}{|c|}{$R=27.5$} &\multicolumn{1}{|c|}{$ R=55 $}&\multicolumn{1}{|c|}{$ R=110 $} &\multicolumn{1}{|c|}{$R=27.5$} &\multicolumn{1}{|c|}{$ R=55 $}&\multicolumn{1}{|c}{$ R=110 $}  \\ 
		\hline
		Standard PS & 7.79 & 31.94&213.35& 0&0&0\\ 
		\hline
		cPS &  8.20 & 32.52 & 224.52& +5.16&+1.83&+5.24\\
		\hline
		DPC & 7.98& 32.59& 227.13&+2.40 &+2.06&+6.46\\
		\hline
	\end{tabular}
	\label{table:DB-ti}
\end{table}

\begin{table}[H]
	\caption{3D dam break: the simulation runtime per physical second denoted as $ t_\mathit{s} $. In this 3D problem, the number of fluid particles would be $ (79,380) $, $ (652,212) $, and $ (5,314,295) $ where $ R=27.5,$ $ 55, $ and $ 110 $, respectively. cPS and DPC reduce the runtime by 2.5-4.5 and 6-8.5 \%, receptively.\strut} 
	\centering
	\small
	\begin{tabular}{p{2.2cm} |c| c| c| c| c| c} 
		\hline
		\multirow{2}{1.8cm}{Model} & \multicolumn{3}{|c|}{$  t_\mathit{s}$ (seconds)} & \multicolumn{3}{|c}{$ (t_\mathit{s}/t^{PS}_\mathit{s}-1)\times100 $}\\
		\cline{2-7} 
		& \multicolumn{1}{|c|}{$R=27.5$} &\multicolumn{1}{|c|}{$ R=55 $}&\multicolumn{1}{|c|}{$ R=110 $} &\multicolumn{1}{|c|}{$R=27.5$} &\multicolumn{1}{|c|}{$ R=55 $}&\multicolumn{1}{|c}{$ R=110 $}  \\ 
		\hline
		Standard PS & 53.76 & 471.84 & 7326.52& 0&0&0\\ 
		\hline
		cPS &  52.38 & 442.88 & 7008.10& -2.58&-6.14&-4.35\\
		\hline
		DPC & 50.57 & 431.54 & 6859.17&-5.93 &-8.54&-6.38\\
		\hline
	\end{tabular}
	\label{table:DB-ts}
\end{table}

\section{Concluding remarks}\label{sec:Conc}
In this study, we introduced effective particle regularization techniques within the framework of the weakly compressible SPH method for the long-term simulation of violent free-surface flows. We represented the dynamic pair-wise particle collision technique (DPC) (originally proposed by \cite{Jandaghian2021_JCP} for the MPS method) and a particle shifting equation coupled with the DPC method (cPS). We implemented these regularization techniques in the DualSPHysics software and simulated four benchmark cases. We validated the numerical simulations with the theoretical solution and the available experimental measurements analyzing the accuracy and convergence of the numerical model. Overall, the qualitative and quantitative results confirmed that the shifting equation (in both the PS and cPS forms) affects the conservation of volume in the long-term simulation of the violent free-surface flows. On the other hand, DPC not only represents more regular particle distribution ensuring numerical stability, but also accurately predicts global evolutions of the system conserving the linear momentum.
 
We summarize the key remarks of this study as follows:
\begin{itemize}
	\item SPH with DPC predicts stable and accurate flow evolution for long-term simulations of free-surface flows.
	\item DPC improves the particle clustering at the free-surface regions and where impact events occur.
	\item DPC respects the linear momentum conservation of the system, while PS affects the global potential energy and unphysically expands the fluid volume.
	\item Numerical results of SPH with DPC converge to the theoretical and expected solutions, however, with PS the convergence behavior of the model would be affected.
	\item The numerical stability achieved by DPC allows the SPH model to adopt larger time steps compared to the PS formulation implemented in DualSPHysics (reducing the simulation runtime by 6-8.5 \%).
\end{itemize}

We should highlight that the more recent improvements to PS formulation (e.g., the works of Jandaghian et al. \cite{Jandaghian2021_JCP} in MPS and Sun et al.\cite{Sun2019} and Hong-Guan and Sun \cite{Hong-Guan2021} in SPH) resolve its numerical issues manifested as the unphysical volume expansion and the particle clustering at the free surface regions. \RvMJ{However, the proposed DPC (as a more effective and efficient alternative technique to the standard PS and the cPS method studied in the present work) is exempted from complex boundary treatments, additional diffusive terms, and their associated computational costs}.

\RvMJ{Future research works may include validating and improving the DPC technique for dealing with the tensile instability issues (i.e., strong negative pressures) in viscous flows with high Reynolds number (like viscous flow past a bluff body). Moreover, it is worthwhile to extend the DPC formulation for simulating high-density ratio multiphase flows.} 
\section*{Acknowledgment}
The authors would like to acknowledge the financially support of the Natural Sciences and Engineering Research Council of Canada (NSERC), Polytechnique Montréal. This study used the high-performance computing resources of Compute Canada and Calcul Quebec.

\appendix
\section{Implementation of DPC in DualSPHysics}
\label{sec:appendixDPC}
\begin{figure}[H]
	\centering
	\includegraphics[width=\textwidth]{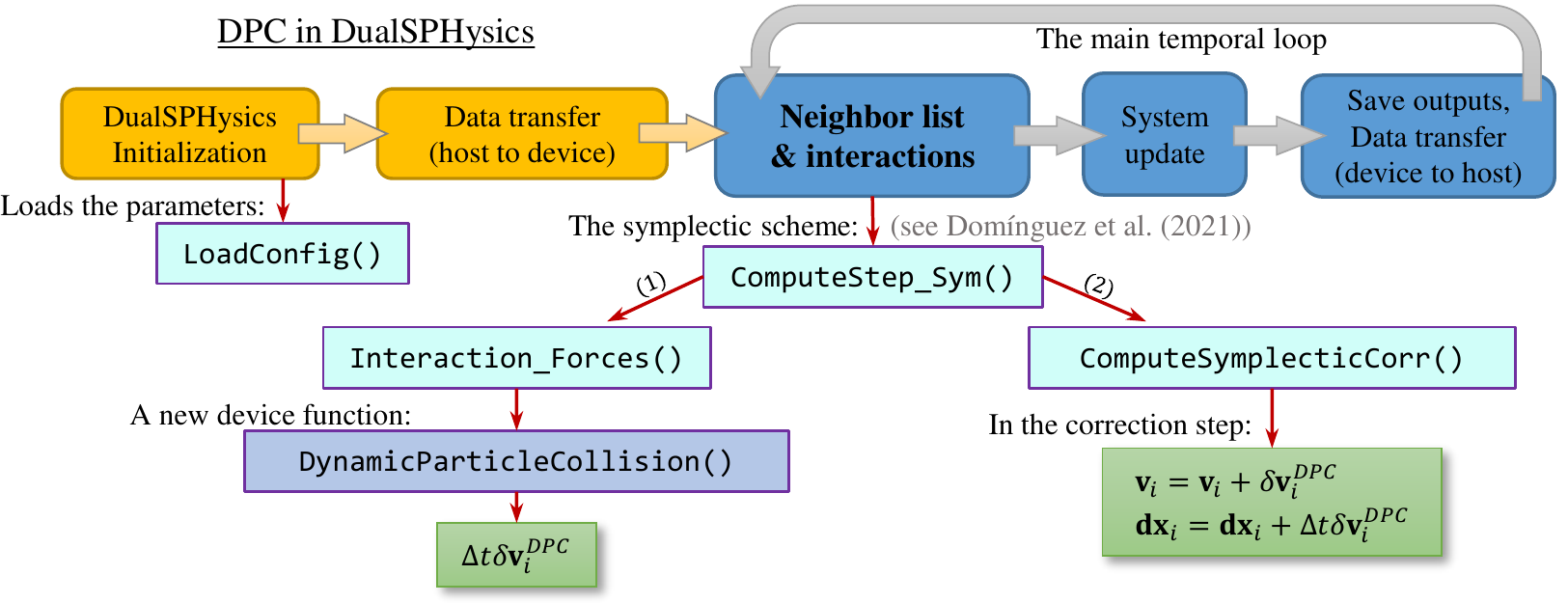}
	\caption{Implementation of the DPC formulation within the GPU-accelerated framework of DualSPHysics. The DPC transport-velocity equation,  $ \delta \mathbf{v}^\mathit{DPC}_i $, is given by (\ref{eq:DPC}).}
	\label{fig_DPC}
\end{figure}

We implement the DPC formulations within the GPU-accelerated subroutines of the DualSPHysics software (v5.0.4). In the initialization stage of the model, the preset parameters of DPC are loaded into the program. In the main temporal loop of calculation, we define a device function, denoted as DynamicParticleCollision(), that implements the DPC term (\ref{eq:DPC}) using the Compute Unified Device Architecture (CUDA) C++ parallel programming language (Fig. \ref{fig_DPC}). This function is nested inside the existing device function that contains the \textit{for loop} responsible for calculating the interaction forces by calling the neighboring particles and loading their variables. The output of the DPC function is $ \Delta t \delta \mathbf{v}^{DPC}_i  $ which updates the velocity and position vectors of the fluid particles in the second stage (i.e., the correction step) of the symplectic algorithm. We changed the present global and device subroutines considering that the developed model can still execute the existing PS method; the user can choose to substitute the PS execution with the DPC technique. One should note that since DPC does not require free-surface detection and is implemented in the current framework of the GPU-accelerated code (without additional \textit{for loop} for the calculations), we expect the efficiency of the application to remain intact.

\section{Supplementary material}
Supplementary material of this article including videos of the simulations can be found online. The DualSPHysics software developed in this study is available from the authors upon reasonable request.
\label{sec:appendixSupp}

\bibliographystyle{elsarticle-num-names} 
\bibliography{References}





\end{document}